\documentclass[%
 reprint,
superscriptaddress,
showkeys,
preprintnumbers,
amsmath,
amssymb,
aps,
prd,
floatfix,
]{revtex4-2}

\usepackage{graphicx}
\usepackage{dcolumn}
\usepackage{bm}
\usepackage[compatibility=false]{caption}
\usepackage{subcaption}
\usepackage{hyperref}

\usepackage{xcolor}
\definecolor{darkblue}{rgb}{0,0,0.5} 

\hypersetup{
    colorlinks=true,
    linkcolor=darkblue,
    citecolor=darkblue,
    urlcolor=darkblue
}
\usepackage{booktabs}

\begin{document}


\title{Negative Masses and Spatial Curvature: Effects on the Neutrino Mass Tension in $\Lambda$CDM and Extended Cosmologies}

\author{Hayyim Pulido-Hernández}
\email{alamhayyim@estudiantes.fisica.unam.mx}
\affiliation{%
Departamento de F\'isica, Instituto Nacional de Investigaciones Nucleares,
Apartado Postal 18-1027, Col. Escand\'on, Ciudad de M\'exico,11801, M\'exico.}
\affiliation{Instituto de Física, Universidad Nacional Autónoma de México, Apartado Postal 20-364, Ciudad Universitaria, Ciudad de México, 04510, México
}

\author{Jorge L.~Cervantes-Cota}
\email{jorge.cervantes@inin.gob.mx}
\affiliation{
Departamento de F\'isica, Instituto Nacional de Investigaciones Nucleares,
Apartado Postal 18-1027, Col. Escand\'on, Ciudad de M\'exico,11801, M\'exico.}

\date{\today}

\begin{abstract}
We investigate the impact of spatial curvature $\Omega_k$ and dynamical dark energy on the cosmological constraints of the neutrino mass sum, $\sum m_\nu$. Using a joint analysis of the latest CMB (Planck and ACT DR6), BAO (DESI DR2) and SNe Ia (DESY5 and DES-Dovekie) datasets, we perform a systematic exploration of the neutrino mass parameter space. To mitigate prior-driven biases near the physical boundary, we implement a symmetric extension wrapper that allows for effective negative masses. We find that the inclusion of spatial curvature modifies the posterior distributions, exhibiting a smooth transition across the $\sum m_\nu = 0$ threshold. In the $\Lambda$CDM + $\Omega_k$ + $\sum m_{\nu,eff}$ framework, we obtain $\sum m_{\nu,eff} = \text{ -} 0.011^{+0.052}_{-0.050}$, reducing the tension with the terrestrial lower limit of $0.06$ eV from $2.59\sigma$ for the $\Lambda$CDM + $\sum m_{\nu,eff}$ model to $1.17\sigma$. For the most flexible scenario $\omega_0\omega_a$CDM + $\Omega_k$ + $\sum m_{\nu,eff}$, we find $\sum m_{\nu,eff} = \text{ -}0.07\pm 0.11$ with a tension of $1.13\sigma$, illustrating how the increased parameter freedom notably degrades the precision of the mass estimate compared to simpler extensions. Our results demonstrate that current cosmological bounds on $\sum m_\nu$ are influenced by boundary effects and geometric degeneracies.
\end{abstract}


\maketitle

\tableofcontents

\begin{section}{Introduction}

The standard model of cosmology assumes that its visible components are 
baryons and photons, for which we have overwhelming direct 
evidence, and it also incorporates dark matter and dark energy (DE), whose gravitational effects we can only perceive indirectly. The final components are 
neutrinos, which are known to possess mass. Recent results from the 
ground-based neutrino oscillation experiments establish, on the one hand, a 
lower bound $\sum m_\nu > 0.059$ eV for normal mass ordering and 
$\sum m_\nu > 0.1$ eV for inverted mass 
ordering \cite{Neutrinos_Terrestres_3}. On the other hand, an upper 
bound comes from the KATRIN experiment, that when combined with 
oscillation results, leads to the limit $\sum m_\nu < 1.35$ eV 
at $90 \%$ cl. \cite{KATRIN:2019yun,KATRIN:2021uub,KATRIN:2024cdt}.

The determination of neutrino mass through cosmological observables has been 
an active and fruitful area of research for decades. Since the first estimate 
in 1966 using astronomical measurements \cite{Gershtein1966}, followed by 
the first measurement using modern methodology in 1999 \cite{Croft_1999}, both 
the precision and volume of available data have increased. Today, 
the availability of large-scale datasets and enhanced computational power allow 
cosmologists to test a wide range of data and model combinations, aiming to 
resolve the inconsistencies currently present in the standard cosmological 
model. Neutrino masses naturally enter the discussion since they induce effects 
in the Hubble expansion rate, CMB power spectra, and clustering properties of 
large scale structure. The measurement of neutrino masses has thus become a 
central objective of major astronomical collaborations. 
\newpage

However, as cosmological data have reached unprecedented precision, a new 
challenge has emerged: the neutrino mass sum inferred from cosmological 
observables is in significant tension with the lower bounds established by 
terrestrial experiments. In fact, posterior probabilities resulting from the 
cosmological inference in the $\Lambda$CDM model seem to prefer negative values. 
This is not a recent artifact, as early results from the Planck collaboration 
already hinted at this trend \cite{Planck:2013nga}, which has only been amplified 
with subsequent data releases \cite{Palanque_Delabrouille_2015, Alam_2021, Aiola_2020, Di_Valentino_2022}. Most recently, the DESI collaboration has 
reported the most stringent constraints to date, which simultaneously represent 
the highest level of tension with local measurements \cite{DESI_DR2_II, DESIDR2_Neutrinos}.

On the other hand, the DESI results also indicate a preference for alternative dark energy models, specifically the Chevallier-Polarski-Linder (CPL, also known as the $\omega_0\omega_a$CDM) parameterization \cite{CPL2001, CPL2003}. This parametrization, however, changes the inferred value of the matter energy density ($\Omega_m$), that in turn affects the neutrino mass. The synergy between DESI's precise $\Omega_m$ measurements and additional datasets, such as CMB, CMB Lensing, and Supernovae type Ia (SN Ia), allows for the breaking of key parameter degeneracies, most notably for the present work, those involving spatial curvature, $\Omega_k$. While the flat-Universe baseline model is standard, treating $\Omega_k$ as a free parameter impacts the total matter density and, consequently, the inferred $\sum m_\nu$. Recent investigations have demonstrated that exploring curved cosmologies can effectively mitigate the tension with terrestrial constraints \cite{itsokcurvature}.

Alongside these developments, this work provides a comprehensive extension to the current literature by analyzing the $\omega_0\omega_a$CDM framework in conjunction with the spatial curvature and also exploring the negative neutrino mass scenario. Although a negative mass is physically nonsensical within the current paradigm, it serves as a valuable phenomenological approach to identify the directions in which current datasets pull the model. To this end, we adopt and numerically implement the methodology proposed in Ref. \cite{negativeneutrinomasses} for an effective neutrino mass sum.

The structure of this work is organized as follows: Section \ref{Sec:NeutrinoEffects} discusses the cosmological role of neutrinos and their interplay with our chosen datasets. Section \ref{Sec: Numerical} details the numerical implementation of the effective neutrino mass and the sampling configuration. Our main findings are presented and discussed in Section \ref{Sec:Results}. In Section \ref{Sec:Conclusions} we summarize our findings and conclude. Finally, Appendix \ref{Apendix: Dovekie_Results} reports relevant results utilizing the most recent supernova recalibration.

\end{section} 

\begin{section}{Neutrino mass effects, cosmic probes, and models}\label{Sec:NeutrinoEffects} 

Although terrestrial experiments can impose upper \cite{KATRIN:2019yun, KATRIN:2021uub, KATRIN:2024cdt} or lower \cite{Neutrinos_Terrestres_1,Neutrinos_Terrestres_2,Neutrinos_Terrestres_3} bounds on the sum of neutrino masses, the most restrictive constraints arise from cosmological measurements \cite{DESIDR2_Neutrinos,Calabrese_2025}. Through the evolution of the Universe, neutrinos are the only particles that transition from being relativistic particles in the early Universe to behaving as cold dark matter in the late Universe. This transition affects the dynamics in various ways, modifying the cosmological observables. The bounds imposed by terrestrial measurements imply that neutrinos begin to behave as dark matter well after the epoch of recombination. Consequently, neutrinos have two primary implications in Large Scale Structure \cite{Loverde_2024}: they modify the geometry of the Universe and suppress the growth of large-scale structure.

When influencing cosmic geometry, neutrinos modify cosmological distances. For the background evolution, a fundamental expression is the Friedmann equation, which takes the form:
\begin{widetext}
\begin{equation}\label{Friedmann}
    \frac{H^2(z)}{H_0^2} = \Omega_r (1+z)^4  + \Omega_m (1+z)^3 + \Omega_k (1+z)^2 + \Omega_{DE}\frac{\rho_{DE}(z)}{\rho_{DE,0}} + \Omega_\nu\frac{\rho_\nu(z)}{\rho_{\nu,0}}
\end{equation}
\end{widetext}
 where the $\Omega_i$ are the known density parameters evaluated at present times, and $\Omega_\nu$ is the neutrino density given by
\begin{equation}\label{neutrino_density}
    \Omega_\nu = \frac{\sum m_\nu}{93.14 h^{2}~eV} \, ,
\end{equation}
where $h$ is given by  $H_0 = 100 \, h $ km s$^{-1}$Mpc$^{-1}$. Furthermore, a key observable quantity is the \textit{transverse comoving distance}, defined as \cite{hogg2000distancemeasurescosmology}:
\begin{equation}\label{Distance_Trans}
    D_M(z) = \begin{cases}
        \frac{1}{H_0\sqrt{\Omega_k}}\sinh\left[\sqrt{\Omega_k}\frac{\chi(z)}{1/H_0}\right] & \text{for }\Omega_k>0\\
        \chi(z) & \text{for }\Omega_k=0\\
        \frac{1}{H_0\sqrt{|\Omega_k|}}\sin\left[\sqrt{\Omega_k}\frac{\chi(z)}{1/H_0}\right] & \text{for }\Omega_k<0
    \end{cases} 
\end{equation}
where the line-of-sight comoving distance is
\begin{equation}\label{Distance_Comoving}
    \chi(z) = \int_{0}^{z}\frac{dz^{\prime}}{H(z^{\prime})} \, .
\end{equation}
All other observable cosmological distances are derived from (\ref{Distance_Comoving}) and the Hubble function, $H(z)$. As we can see from equations (\ref{Friedmann}-\ref{Distance_Trans}), the sum of neutrino masses affects the cosmic expansion history. This primarily affects low-redshift observations, and although the effect is tiny, a key point is the precise determination of $\Omega_m$ in equation (\ref{Friedmann}), since it conditions the influence of the neutrino mass that, in late times, behaves as dark matter.

\newpage

The primary probes for these distance measurements are Baryon Acoustic Oscillations (BAO) and Type Ia Supernovae (SNe Ia), while Cosmic Microwave Background (CMB) data provide a precise determination of the matter density through the full temperature and polarization power spectra and through the angular scale of the sound horizon, $\theta_{*}$. The tight constraints on $\theta_{*}$ provided by the CMB induce a geometric degeneracy between $\Omega_m$ and $\sum m_\nu$. Following the arguments in Ref. \cite{Loverde_2024}, maintaining the observed angular scale requires $\Omega_m$ to adjust in response to changes in the sum of the neutrino masses.  

On the other hand, there is a suppression of the large-scale structure caused by massive neutrinos. Although at late times neutrinos behave as dark matter, they retain thermal velocities high enough to inhibit the growth of structure on small scales. While neutrinos are relativistic, their free-streaming length, $x_{fs}$, is of the order of the Hubble radius. As the Universe evolves, this length decreases, and the free-streaming scale defined as $k_{fs} = 1/x_{fs}$ takes the form \cite{Neutrino_FS}:
\begin{equation}\label{Free-Streaming}
    k_{fs}(z) \thickapprox 0.0908 \frac{H(z)}{(1+z)^2}\left( \frac{ m_\nu}{0.1eV} \right) hMpc^{-1} .
\end{equation}
The streaming effect applies to every neutrino mass that, for simplicity, we assume here is degenerate.
Any matter overdensity with a radius smaller than the free-streaming length will be attenuated. This suppression of large-scale structure primarily affects the matter power spectrum, although it also has detectable consequences in the CMB \cite{Lesgourgues_2012,Loverde_2024}. To characterize the cutoff in the matter spectrum induced by free-streaming, a Full-Shape analysis can be employed. Given the precision of the datasets, it has not yet been detected; therefore, we will not employ a Full-Shape analysis here.

On the other hand, an important perturbative implication of neutrino 
mass originates when CMB photons travel from the surface of last 
scattering toward us, and their trajectories are deflected by 
the gravitational potential of the large-scale structures along 
their path. This phenomenon, known as CMB lensing, acts as a 
probe of the matter distribution to which neutrinos also 
contribute, allowing for the extraction of indirect but 
robust information regarding their masses. That is why, in 
this analysis, CMB lensing is also included. 

Consequently, it is imperative to combine multiple observables such as BAO, 
SNe Ia, CMB, and CMB Lensing to break the parameter degeneracies inherent in 
each individual survey. However, a notable challenge arises when 
utilizing the latest datasets, as their respective estimates for $\Omega_m$ 
show mutual inconsistency in the $\Lambda$CDM model. Specifically, DESI BAO 
measurements report low values ($\Omega_m=0.2975\pm0.0086$ 
\cite{DESI_DR2_II}), while Planck CMB data favor intermediate results 
($\Omega_m = 0.3166\pm0.0084$ \cite{Planck2018_Results}), and supernova 
surveys suggest even higher densities ($\Omega_m>0.35$) 
\cite{descollaboration2024, PantheonPlus, Union3}. The discrepancy among 
these datasets creates a tension in the determination of the expansion 
history. 

\bigskip

Consequently, to reconcile the matter density value, the 
statistical fit is driven towards an effective negative neutrino mass as a 
compensatory mechanism \cite{Loverde_2024}.

Another relevant extension to this analysis is the study of spatial curvature. While a spatially flat Universe ($\Omega_k=0$) is conventionally assumed, CMB measurements exhibit a significant geometrical degeneracy between curvature and the matter density $\Omega_m$. Breaking this degeneracy requires the inclusion of other probes, such as supernovae or BAO, which provide robust constraints on $\Omega_m$. This interplay directly affects neutrino mass estimates, as neutrinos contribute to the total matter density at late times. Ref. \cite{itsokcurvature} provides a thorough exploration of these physical effects within the $\Lambda$CDM model, demonstrating that allowing for non-zero spatial curvature can alleviate the tension between cosmological bounds and terrestrial neutrino mass measurements. In this work, we consider curvature in our analysis.

On the theoretical side, the matter and neutrino density parameters, as seen in Eq. (\ref{Friedmann}), depend on the DE model. To explore extensions beyond the $\Lambda$CDM model, one may adopt the CPL parametrization, also referred to as $\omega_0\omega_a$CDM. In this framework, the evolution of the dark energy density in (\ref{Friedmann}) is governed by a time-dependent equation-of-state parameter, defined as:
\begin{equation}\label{Eq:CPL_Parameter}
    \omega(z) = \omega_0 + \omega_a \frac{z}{1+z}
\end{equation}
where $\omega_0$ and $\omega_a$ are the parameters to be determined. One normally expects that parameter inferences should yield results in which $\omega_0 =-1$ and $\omega_a =0$ are within error bars that include the $\Lambda$CDM model. However, recent results from DESI have shown a tension of 1.7-4.2$\sigma$ relative to $\Lambda$CDM, depending on the specific combination of datasets employed \cite{DESI_DR2_II}; see, however, Ref. \cite{Chudaykin:2026nls} for a reanalysis finding a lower tension. This discrepancy motivates the exploration of scenarios involving dynamical dark energy. Indeed, recent independent analyses have further bolstered this motivation, reporting a $5\sigma$ statistical preference for dynamical dark energy over the standard $\Lambda$CDM baseline \cite{Challenging-LCDM}. Furthermore, evaluating these dynamical scenarios in tandem with spatial curvature, such as in \cite{101093/mnrasl/slaf063}, is proving to be a highly relevant approach to unraveling the complex degeneracies inherent in modern cosmological datasets.

To illustrate the current observational landscape, Table \ref{tab:art_neutrinos} provides a detailed compendium of reported bounds on $\sum m_\nu$. The results are presented as a function of the various cosmological models and data combinations employed in each study, allowing for the identification of how these constraints vary across different assumptions and observational samples. This comparison facilitates an analysis of the impact that both model choice and data selection have on the precision of the resulting limits. 

\begin{table*}[!t]
\begin{ruledtabular}
    \centering 
        \resizebox{0.8\textwidth}{!}{%
            \begin{tabular}{l c}
                \toprule
                \textbf{Dataset} & $\mathbf{\sum m_\nu}$ [eV] (95\%) \\
                \midrule
                \multicolumn{2}{c}{\textbf{\color{blue} $\Lambda$CDM Model}} \\
                \midrule
                
                DESI DR2+CMB & $< 0.0642$ \cite{DESIDR2_Neutrinos} \\
                DESI DR2+CMB+DESY5 & $< 0.0744$ \cite{DESIDR2_Neutrinos}\\
                DESI DR2+CMB+Union3 & $< 0.0674$ \cite{DESIDR2_Neutrinos}\\
                DESI DR2+CMB+Pantheon+ & $< 0.0704$ \cite{DESIDR2_Neutrinos}\\
                DESIDR1(FS+BAO)+BBN+$n_s$ & $< 0.300$ \cite{DESIDR2_Neutrinos}\\
                DESIDR1(FS+BAO)+BBN+$(\theta_{*}, n_s)_{CMB}$ & $< 0.193$ \cite{DESIDR2_Neutrinos}\\
                DESI DR2+CMB+DESY5+DESY1(WL) & $< 0.085$ \cite{DESY1Anal}\\
                DESI DR2+CMB+Lensing+DESY5+DESY1(WL) & $0.19^{+0.15}_{0.18}$ \cite{Roy_Choudhury_2025}\\
                
                \midrule 
                
                \multicolumn{2}{c}{\textbf{\color{blue} $\omega$CDM Model}} \\
                \midrule
                
                DESI DR2 + CMB & $< 0.0851$ \cite{DESIDR2_Neutrinos} \\
                DESI DR2 + CMB + DESY5 & $< 0.0586$ \cite{DESIDR2_Neutrinos} \\
                DESI DR2 + CMB + Union3 & $< 0.0649$ \cite{DESIDR2_Neutrinos} \\
                DESI DR2 + CMB + Pantheon+ & $< 0.0653$ \cite{DESIDR2_Neutrinos} \\
                DESI DR2+CMB+DESY5+DESY1(WL) & $< 0.065$ \cite{DESY1Anal} \\
                
                \midrule
                
                \multicolumn{2}{c}{\textbf{\color{blue} $\omega_0\omega_a$CDM Model}} \\
                \midrule
                
                DESI DR2 + CMB & $< 0.163$ \cite{DESIDR2_Neutrinos} \\
                DESI DR2 + CMB + DESY5 & $< 0.129$ \cite{DESIDR2_Neutrinos} \\
                DESI DR2 + CMB + Union3 & $< 0.139$ \cite{DESIDR2_Neutrinos} \\
                DESI DR2 + CMB + Pantheon+ & $< 0.117$ \cite{DESIDR2_Neutrinos} \\
                DESI DR2+CMB+DESY5+DESY1(WL) & $ 0.098^{+0.016}_{-0.037}$ \cite{DESY1Anal} \\
                \toprule
                
                \multicolumn{2}{c}{\textbf{\color{blue} \textit{Matter Conversion} Model}} \\
                \midrule
                
                DESI DR2 + CMB(No Lensing) & $ 0.106^{+0.050}_{-0.069}$ \cite{AA} \\
                
                \midrule 
                
                \multicolumn{2}{c}{\textbf{\color{blue} EDE Model}} \\
                \midrule
                
                DESIDR1+CMB  & $<0.097$ \cite{axion} \\
                
                \midrule
                
                \multicolumn{2}{c}{\textbf{\color{blue} Sound Horizon Agnostics}} \\
                \midrule
                
                DESI DR2+Lensing+Pantheon++DESY3(WL) & $0.55^{+0.23}_{-0.37}$ \cite{Helena} \\ 
                IDL+unSNeIa+KiDS+BBN+$A_s$+$n_s$ & $1.08^{+0.65}_{-0.60}$ \cite{Lesg} \\
                IDL+unSNeIa+KiDS+BBN++$A_s$+$n_s$+Lensing(No SPT-3G) & $1.01^{+0.52}_{-0.46}$ \cite{Lesg} \\

                \midrule 
                
                \multicolumn{2}{c}{\textbf{\color{blue} $\tau$ modification}} \\
                \midrule
                
                DESIDR1+CMB(NoLow$\ell$, NoLensing)  & $0.01 \pm 0.06$ \cite{dispu_tau_able} \\

                \midrule 
                
                \multicolumn{2}{c}{\textbf{\color{blue} Minimally modified gravity}} \\
                \midrule
                
                DESIDR2+CMB  & $< 0.135$ \cite{q1ng-2xvp} \\

                \midrule 
                
                \multicolumn{2}{c}{\textbf{\color{blue} Neutrino decays}} \\
                \midrule
                
                DESIDR2+CMB  & $< 0.23$ \cite{lwdv-qrcc} \\

                \bottomrule
            \end{tabular}%
        }
        \caption{Compendium of recent estimates for $\sum m_\nu$ across various cosmological models and datasets. Although the DESI collaboration \cite{DESIDR2_Neutrinos} provides the most extensive set of measurements, their results for the negative mass regime are omitted in this table for brevity. Notably, Ref. \cite{DESY1Anal} establishes a Gaussian constraint within the positive mass range using weak lensing data. Finally, recent studies \cite{AA, dispu_tau_able} illustrate the diverse strategies for addressing the neutrino mass tension; there is a growing trend toward modifying early-Universe physics.}
    \label{tab:art_neutrinos}
    \end{ruledtabular}
    \end{table*}

\end{section} 

\begin{section}{Numerical Implementation and datasets}\label{Sec: Numerical}

Our parameter estimation is performed via Markov Chain Monte Carlo (MCMC) 
sampling, specifically employing the Metropolis-Hastings algorithm within 
the cosmological inference 
code Cobaya \cite{Torrado_2021_COBAYA, 2019ascl.soft10019T_COBAYA}. Our 
analysis utilizes the Boltzmann solver CAMB \cite{Lewis_2000_CAMB} 
for positive neutrino mass priors and CLASS \cite{CLASS_II, CLASS_IV} for 
the runs that include the negative mass regime, as well as Cobaya's public
likelihoods. We run four independent chains in parallel for each case. Furthermore, point 
estimates for the best-fit parameters are obtained using CLASS with the \texttt{iminuit}
minimizer integrated within Cobaya. The convergence of the MCMC is determined 
by the Gelman-Rubin statistics \cite{Gelman_Rubin}, requiring $R-1<0.01$. 

Post-processing 
is done using the \texttt{GetDist} Python package \cite{lewis2019_getdist}.

\vspace{-15pt}

\begin{subsection}{Effective neutrino mass}

\vspace{-12pt}

As mentioned, Table \ref{tab:art_neutrinos} illustrates recent findings reported within the standard 
framework in which neutrino masses are always positive. While a negative mass for a particle lacks any 
physical meaning, as we discussed in Section \ref{Sec:NeutrinoEffects}, various results exhibit a tension 
with the limits established by terrestrial experiments and even their posterior probabilities lie in the 
negative values, when one allows for it  
\cite{DESIDR2_Neutrinos, dispu_tau_able,craig2024nusgoodnews,2019ascl.soft10019T_COBAYA,naredotuero2024livingedgecriticallook}. To find out whether this is really the 
case, that data prefer negative masses, a methodology has recently been put forward 
\cite{negativeneutrinomasses}. In that work, two approaches were presented. The first, named the 
\textit{exact method}, consists of modifying the background and perturbative equations as follows

\begin{widetext}
\begin{equation}
    \begin{split}
        \epsilon_{eff} = sgn&(m_\nu)\sqrt{x^2+a^2m_\nu^2/T_\nu^2}\\
        \dot{\mathrm{\Psi}}_0 = \text{ -}\frac{qk}{\epsilon_{eff}}\mathrm{\Psi}_1 \text{ - } \dot{\phi}\frac{d\ln{\bar{f}}}{d\ln{q}}, \hspace{0.5cm} \dot{\mathrm{\Psi}}_1 &= \frac{qk}{3\epsilon_{eff}}\left( \mathrm{\Psi}_0 \text{ - } 2\mathrm{\Psi}_2 \right) \text{ - } \frac{\epsilon_{eff}k}{3q}\psi\frac{d\ln{\bar{f}}}{d\ln{q}} \\
        \dot{\mathrm{\Psi}}_\ell = \frac{qk}{(2\ell+1)\epsilon_{eff}}&\left[ \ell\mathrm{\Psi}_{\ell-1} \text{ - } (\ell+1)\mathrm{\Psi}_{\ell+1} \right], (\ell\geq 2)
    \end{split}
\end{equation}
\end{widetext}
where $\epsilon_{eff}$ is the effective neutrino energy parameter and $\Psi_\ell$ are the distribution function perturbation multipole moments in the Newtonian gauge \cite{Ma:1995ey}.

The second approach is the \textit{extrapolation method}, which requires modifying a physical observable $X$ that depends on a set of parameters $\theta$.

\begin{widetext}
\begin{equation}\label{Eq:Mnu_eff}
    X_{\theta}^{\sum m_{\nu,eff}} \equiv X_{\theta}^{\sum m_{\nu}=0} + sgn\left(\sum m_{\nu,eff}\right) \left[ X_{\theta}^{|\sum m_{\nu,eff}|} \text{ - } X_{\theta}^{\sum m_{\nu}=0}  \right] \, .
\end{equation}

\end{widetext}

\noindent The first method requires a modification of the cosmo code in CAMB or CLASS, whereas the second method demands modifying the interface between CLASS and Cobaya. Since both methods yield the same results, we thus chose the extrapolation method in this work.

The implementation requires determining the value of the \texttt{m\_ncdm} variable at each step of the Cobaya sampler; this variable defines the neutrino mass within CLASS. Two scenarios then arise:\begin{itemize}
        \item[A.] If \texttt{m\_ncdm}$\geq$0: Cobaya and CLASS execute normally.
        \item[B.] If \texttt{m\_ncdm}$<$0: CLASS is executed twice to obtain $X_{\theta}^{|\sum m_{\nu,eff}|} $ and $ X_{\theta}^{\sum m_{\nu}=0}$ for each observable required by Cobaya, followed by the application of Eq. (\ref{Eq:Mnu_eff}).
    \end{itemize}
Figure \ref{fig:Validation} displays the results 
obtained to verify the correct functioning of the code. In the upper panel, we 
confirm that the direct application of extrapolation method within CLASS is the 
same as the computation performed via Cobaya. Furthermore, the lower panel 
shows an agreement of our results and those presented in \cite{DESIDR2_Neutrinos}. 

\begin{figure}[!t]
     \centering
     \begin{subfigure}[b]{0.4\textwidth}
         \centering
         \includegraphics[width=\textwidth]{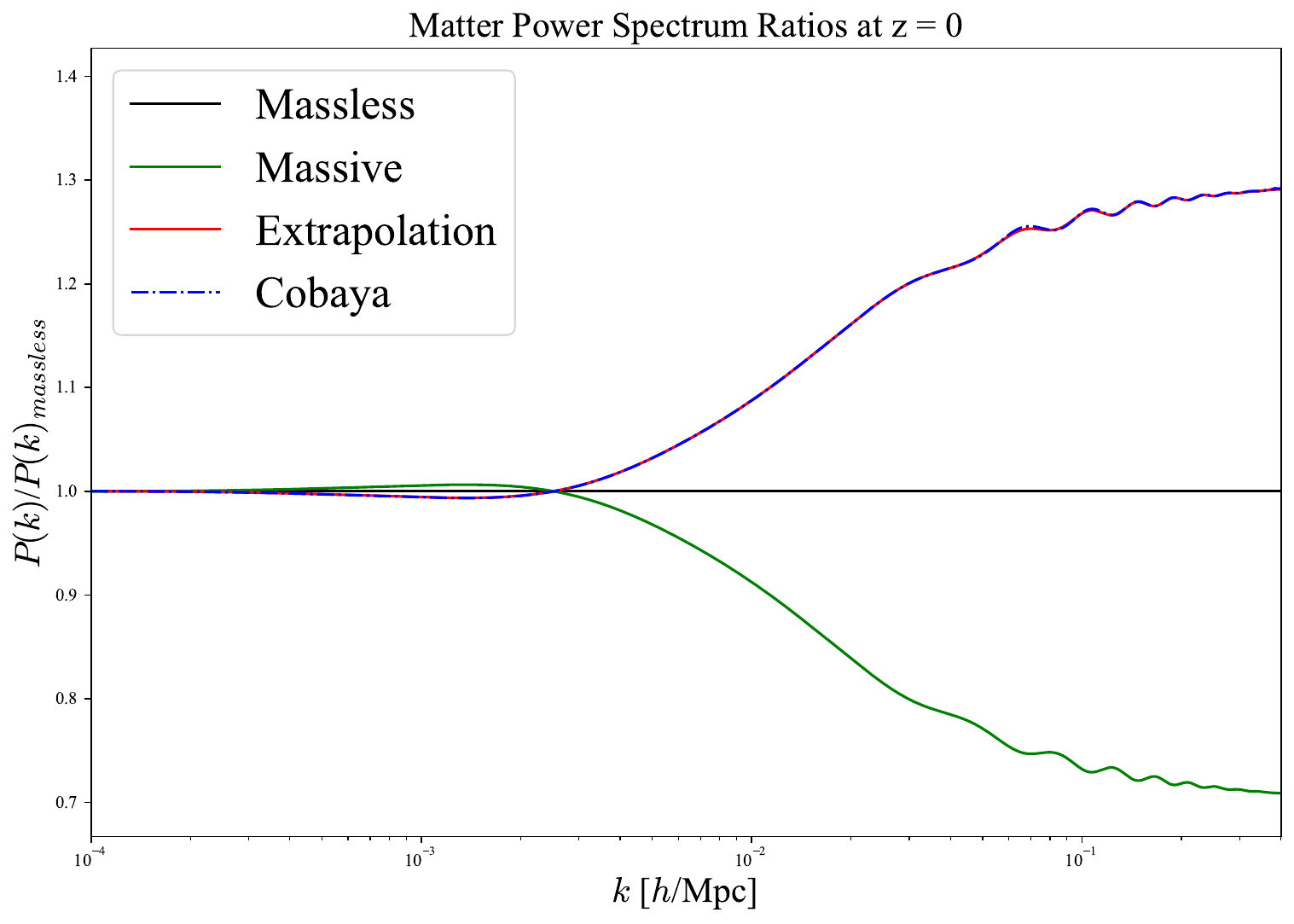}
     \end{subfigure}
     \hfill
     \begin{subfigure}[b]{0.45\textwidth}
         \centering
         \includegraphics[width=\textwidth]{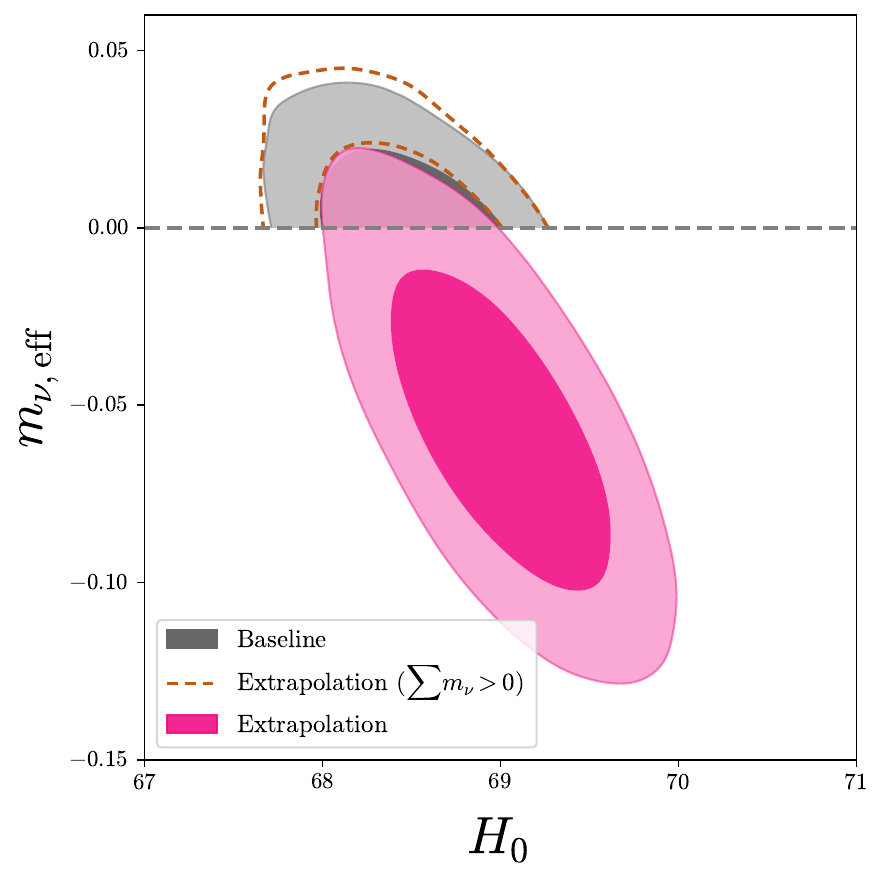}
     \end{subfigure}
     \caption{Validation of the code implementation for the extrapolation method. \textbf{(Upper)} Comparison of ratios between negative mass scenarios and the massless case; the red line represents the direct application of Eq. (\ref{Eq:Mnu_eff}) within CLASS, while the blue line corresponds to the computation performed via the Cobaya \texttt{get\_model()} function. \textbf{(Lower)} Posterior contours comparing the baseline case with a positive mass prior, the extrapolation framework restricted to a positive prior, and the full implementation utilizing a negative prior.}
     \label{fig:Validation}
\end{figure}
    
\end{subsection}

\begin{subsection}{Datasets for the analysis}\label{Sec:Datasets}

 Our objective is to obtain constraints on $\sum m_{\nu}$, considering positive and 
 negative priors on it, and to understand the role played by curvature and the dark 
 energy parameters $\omega_0$ and $\omega_a$ of the CPL parameterization. For this purpose, we 
 have employed the subsequent datasets, with our baseline configuration including:

 \vspace{70pt}

\begin{itemize}
    \item \textbf{BAO}: We use BAO data from DESI DR2 \cite{DESI_DR2_II} using all of the tracers.
    \item \textbf{CMB}: We use the Planck 2018 \cite{Planck2018_Results} PR3 low-$\ell$ for TT and EE modes and PR4 high-$\ell$ CamSpec \cite{CamSpec_1, Tristram_2021, Tristram_2024_PR4, Efstathiou_2021_CamSpecPipeline} TTTEEE likelihoods. Additionally, we use the combination of Planck and ACT DR6 CMB lensing \cite{Madhavacheril_2024_ACTDR6,ACT_DR6_Lensing,Carron_2022_ACT+Planck}.
    \item \textbf{SNe Ia}: We use the DESY5 \cite{descollaboration2024} likelihood for supernovae measurements.
\end{itemize}

These specific data, when analyzed individually, imply a tension in the inferred $\Lambda$CDM cosmological parameters, e.g. in $\Omega_m$. While this 
would, in principle, be a reason to be cautious in its use, we have chosen 
this data setup for two primary reasons: first, to examine the behavior of parameter estimates under $\Lambda$CDM extensions and alternative models; and second, to ensure the inclusion of the most recent data releases, omitting the Planck LolliPop and HiLLiPoP likelihoods as CamSpec likelihoods yields equivalent results \cite{Jense_2026} while reducing the overall computational time. 

\bigskip

Recently, a recalibration of the DESY5 data was released, designated as DESY5-Dovekie \cite{popovic2025darkenergysurveysupernova}. We have verified that the use of the updated sample does not alter our primary findings regarding the sum of neutrino masses. Consequently, the analysis involving the Dovekie recalibration is presented in Appendix \ref{Apendix: Dovekie_Results}.

Our runs make use of the set of seven cosmological parameters ($100\theta_{*}$, $\ln(10^{10}A_s)$, $n_s$, $\tau$, $\omega_{cdm}$, $\omega_{b}$, $\Omega_k$), where $\theta_{*}$ is the acoustic angular scale, $\ln(10^{10}A_s)$ and $n_s$ 
are the amplitude and spectral index of the primordial scalar perturbations, respectively, $\tau$ is the reionization optical depth, $\omega_{cdm}$ and $\omega_{b}$ are the physical abundance of cold dark matter and baryons, respectively, and $\Omega_k$ is the spatial curvature parameter. 

\newpage

When we analyze the case of a dynamical DE we use the CPL parametrization that 
adds two extra parameters, $w_0$ and $w_a$. For each DE model, one run was performed with fixed neutrino mass and another with varying neutrino mass. When 
$\sum m_\nu$ is sampled, we assume three degenerate mass eigenstates while in the opposite case $\sum m_\nu$ is fixed to a value of 0.06 eV assuming a single non-zero mass neutrino.

In Table \ref{tab: Cobaya_Priors} are presented the priors used for each parameter, following those of the DESI collaboration \cite{DESIDR2_Neutrinos}. 

\begin{table}[!t]
\centering
\resizebox{0.45\textwidth}{!}{%
\begin{tabular}{|lll|}
\hline
Model & Parameter & Prior \\ \hline
 \textbf{$\Lambda$CDM}     &   $\omega_{cdm}$        &   $\mathcal{U}[0.001, 0.99]$    \\
      &     $\omega_{b}$      &     $\mathcal{U}[0.005, 0.1]$   \\
      &     $100 \theta_{*}$      &     $\mathcal{U}[0.5, 10]$   \\
      &     $\ln(10^{10}A_s)$      &   $\mathcal{U}[1.61, 3.91]$     \\
      &      $n_s$     &    $\mathcal{U}[0.8, 1.2]$    \\
      &      $\Omega_k$     &    $\mathcal{U}[\text{ -}0.3, 0.3]$    \\
      &        $\tau$   &    $\mathcal{U}[0.01, 0.8]$    \\ \hline
 \textbf{CPL DE}     &     $\omega_0$      &      $\mathcal{U}[\text{ -}3, 1]$  \\
      &      $\omega_a$     &     $\mathcal{U}[\text{ -}3, 2]$   \\ \hline
   \textbf{Adding Neutrinos}   &    $\sum m_\nu$        &     $\mathcal{U}[0, 5]$  \\
      &    $\sum m_{\nu,eff}$       &     $\mathcal{U}[\text{ -}5, 5]$   \\ \hline
\end{tabular}
}
\caption{Set of parameters and their respective priors used in the analysis. Notice that all priors are flat priors in order to let MCMC explore an extensive configuration space for the models.}
\label{tab: Cobaya_Priors}
\end{table}

\end{subsection}

\end{section}

\vspace{-15pt}
\begin{section}{Results}\label{Sec:Results}

This section presents the results obtained from the cosmological inference process. The discussion is organized into three subsections: Section \ref{Sec:LCDM_Results} details the findings for the standard $\Lambda$CDM model, assuming a positive neutrino mass prior; the same goes for Section \ref{Sec:CPL_Results}, but here the CPL dynamical dark energy model is adopted. Section \ref{Sec:negnu_Results} provides the outcomes obtained when the negative neutrino mass regime is explored for both DE models. Additionally, Section \ref{Sec:Fit_Results} presents various goodness-of-fit criteria obtained via \texttt{iminuit}, which is used to evaluate model performance and to determine the statistical preference for each model.

\begin{subsection}{\texorpdfstring{$\Lambda$}{L}CDM model}\label{Sec:LCDM_Results}

First, we present the results for the $\Lambda$CDM model. Figure \ref{fig:LCDM_Pos_Tri} displays the 2D posterior contours for selected parameters, comparing the $\Lambda$CDM model in both spatially flat
and non-flat scenarios. As illustrated, the inclusion of spatial curvature shifts $\sum m_{\nu}$ to larger positive values, improving the constraints on the neutrino mass sum, a result consistent with the findings in Ref. \cite{itsokcurvature}. When $\Omega_k$ is treated as a free parameter, the resulting bound is
\begin{equation}
    \text{$\Lambda$CDM+$\Omega_k$ : } \sum m_\nu < 0.135 \text{ eV} \hspace{6pt}(95\%) \, .
\end{equation}
As expected, allowing $\Omega_k$ to vary impacts the estimation of $\Omega_m$ which in turn indirectly influences the neutrino mass constraints due to the geometric degeneracy. Regarding the spatial curvature, while the neutrino mass is fixed, we find a $2.15\sigma$ deviation from a flat Universe; we quantify the statistical tensions using a standard Gaussian metric. However, when the neutrino mass is also varied, this statistical significance relaxes to $1.65\sigma$ from the flat model. 

\begin{figure}[t]
        \centering
        \includegraphics[width=\linewidth]{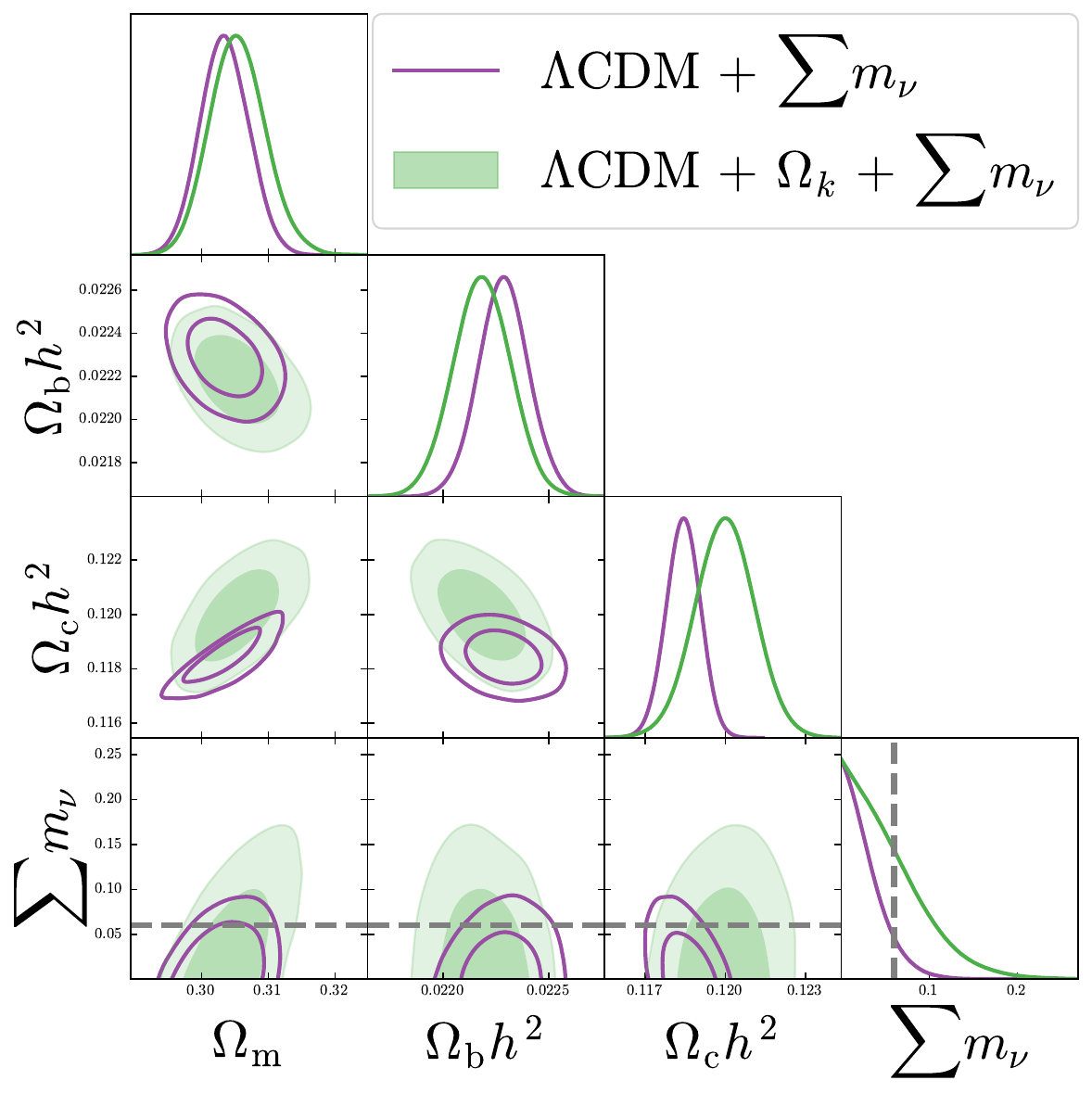}
        \caption{Triangular plot showing the confidence contours for the $\Lambda$CDM+$\sum m_\nu$ (purple) versus the $\Lambda$CDM+$\Omega_k$+$\sum m_\nu$ (green) models. As observed, including curvature as a varying parameter affects $\Omega_m$, causing the associated densities $\Omega_b$, $\Omega_{cdm}$ and $\Omega_\nu$ to be affected as well. The grey dashed line corresponds to the lower bound value $\sum m_\nu = 0.06$ eV.}
        \label{fig:LCDM_Pos_Tri}
\end{figure}

\end{subsection}

\begin{subsection}{\texorpdfstring{$\omega_0\omega_a$}{w0wa}CDM Model}\label{Sec:CPL_Results}

Motivated by the recent DESI results \cite{descollaboration2024,DESI_DR2_II,DESI:2025fii,DESI:2025wyn} favoring a dynamical dark energy scenario, we extend our analysis to include the $\omega_0\omega_a$CDM model. In this scenario, the difference between the flat Universe case and the one with curvature is not as noticeable as it is in the $\Lambda$CDM model, as we can see in the Figure \ref{fig:CPL_Pos_tri}. Even so, the neutrino mass constraint is relaxed relative to the flat case, yielding a value of

\begin{equation}
    \text{$\omega_0\omega_a$CDM+$\Omega_k$ : } \sum m_\nu < 0.171 \text{ eV} \hspace{6pt}(95\%)
\end{equation}

In this case, the presence of curvature affects the $\omega_0$ and $\omega_a$ parameters; as seen in Figure \ref{fig:CPL_Con_w0wa}. It is clear that curvature reduces the tension found in the $\Lambda$CDM model (this being the case for $\omega_0 = \text{ -}1$ and $\omega_a=0$).

\begin{figure}[!t]
    \centering
    \includegraphics[width=\linewidth]{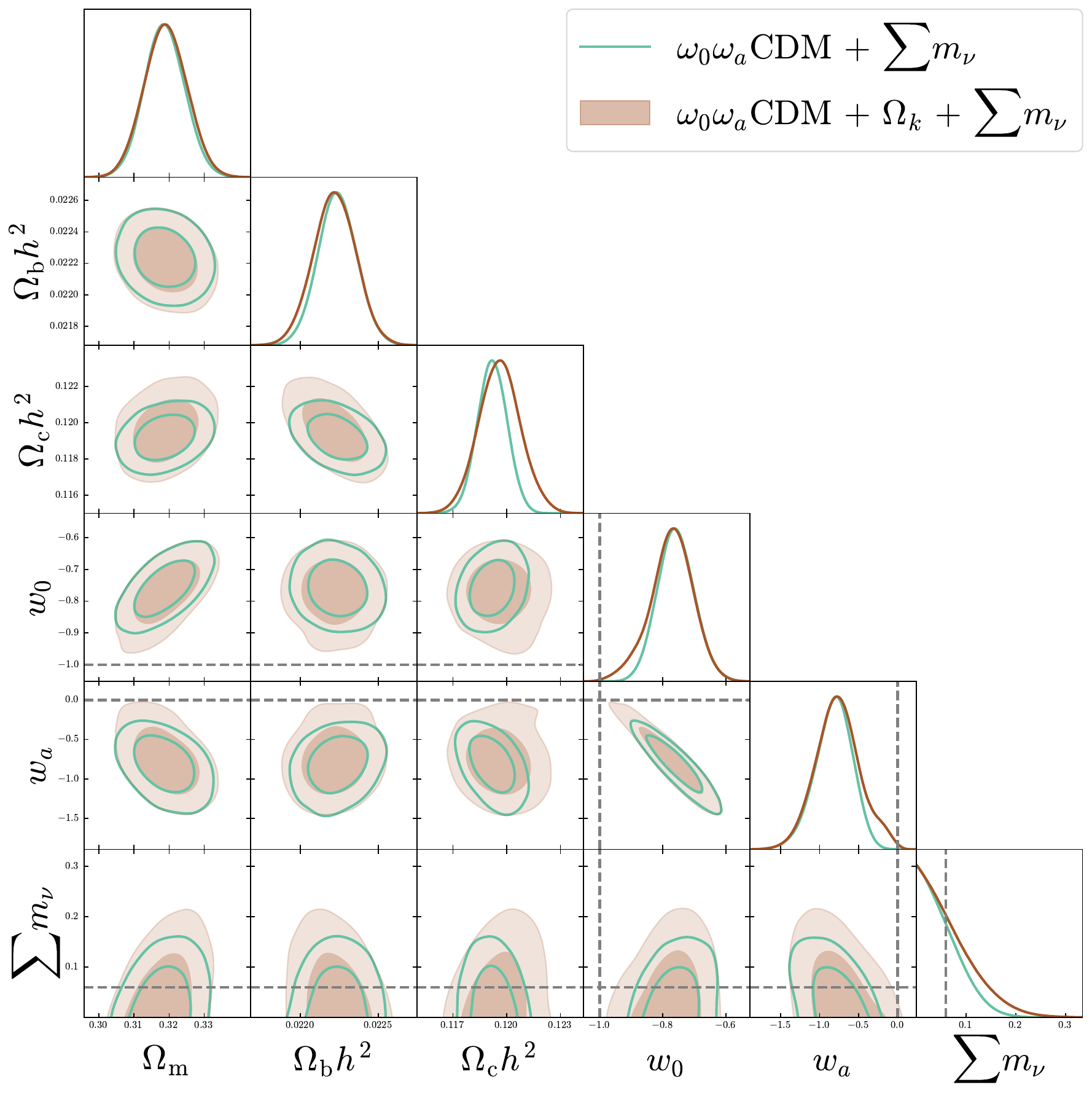}
    \caption{Triangular plot showing the confidence contours for $\omega_0\omega_a$CDM+$\sum m_\nu$ (lime) against $\omega_0\omega_a$CDM+$\Omega_k$+$\sum m_\nu$ (brown). No important shift is observed in the $\Omega_m$ posterior, whereas $\Omega_{c}$ and $\sum m_\nu$ show a slight sensitivity to the inclusion of curvature, with the effect being more pronounced for the former.} 
    \label{fig:CPL_Pos_tri}
\end{figure}

\begin{figure}[!t]
    \centering
    \includegraphics[width=0.8\linewidth]{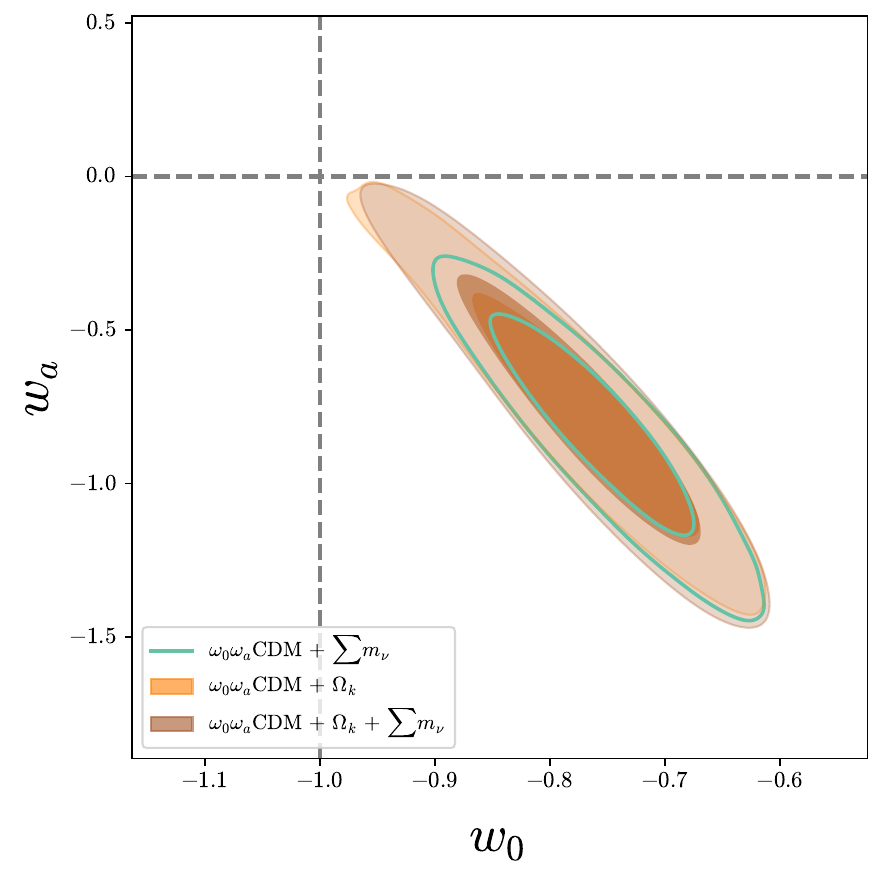}
    \caption{2D plot of the $\omega_0$ and $\omega_a$ parameters. It can be observed that the contours for the $\omega_0\omega_a$CDM+$\Omega_k$ and $\omega_0\omega_a$CDM+$\Omega_k$+$\sum m_\nu$ models are nearly identical, while showing an important difference from $\omega_0\omega_a$CDM+$\sum m_\nu$; this clearly indicates that the shift is driven entirely by curvature. However, the $68\%$ C.L. contour is quite similar across the three models, with the statistical deviation becoming apparent at the $95\%$ C.L., as clearly shown in Figure \ref{fig:CPL_Pos_tri}.}
    \label{fig:CPL_Con_w0wa}
\end{figure}
    
\end{subsection}

\newpage

The next step is to compare the $\Lambda$CDM results with those of the $\omega_0\omega_a$CDM model. Figure \ref{fig:1D_Pos_Posteriors} shows the posterior distributions for $\sum m_\nu$ and $\Omega_k$. Note that all posteriors of $\sum m_\nu$ (upper panel) tend to peak near its lower limit, suggesting that they lie at negative values; however, the presence of curvature seems to alleviate this trend. Also interesting in the upper panel is to note that the posteriors for $\Lambda$CDM + $\Omega_k$ + $\sum m_\nu$ and $\omega_0\omega_a$CDM + $\sum m_\nu$ are nearly identical, demonstrating that the former achieves the same improvement as the latter with one fewer parameter. Concerning $\Omega_k$ (right panel), both models exhibit a preference for a positive value, echoing the findings in \cite{itsokcurvature}, which show that the relationship between $\sum m_\nu$ and $\Omega_k$ is not bilateral, as curvature affects the neutrino mass, but the reverse is not true in the same proportion.

\begin{figure}[!t]
     \centering
     \begin{subfigure}[b]{0.49\textwidth}
         \centering
         \includegraphics[width=\textwidth]{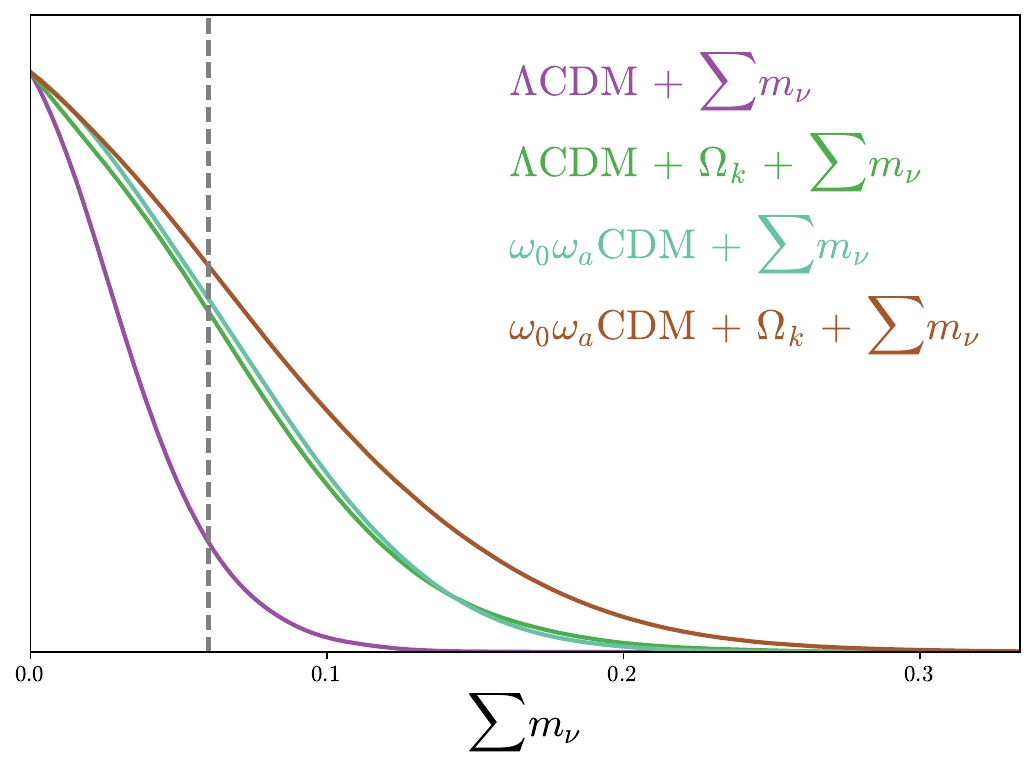}
     \end{subfigure}
     \hfill 
     \begin{subfigure}[b]{0.49\textwidth}
         \centering
         \includegraphics[width=\textwidth]{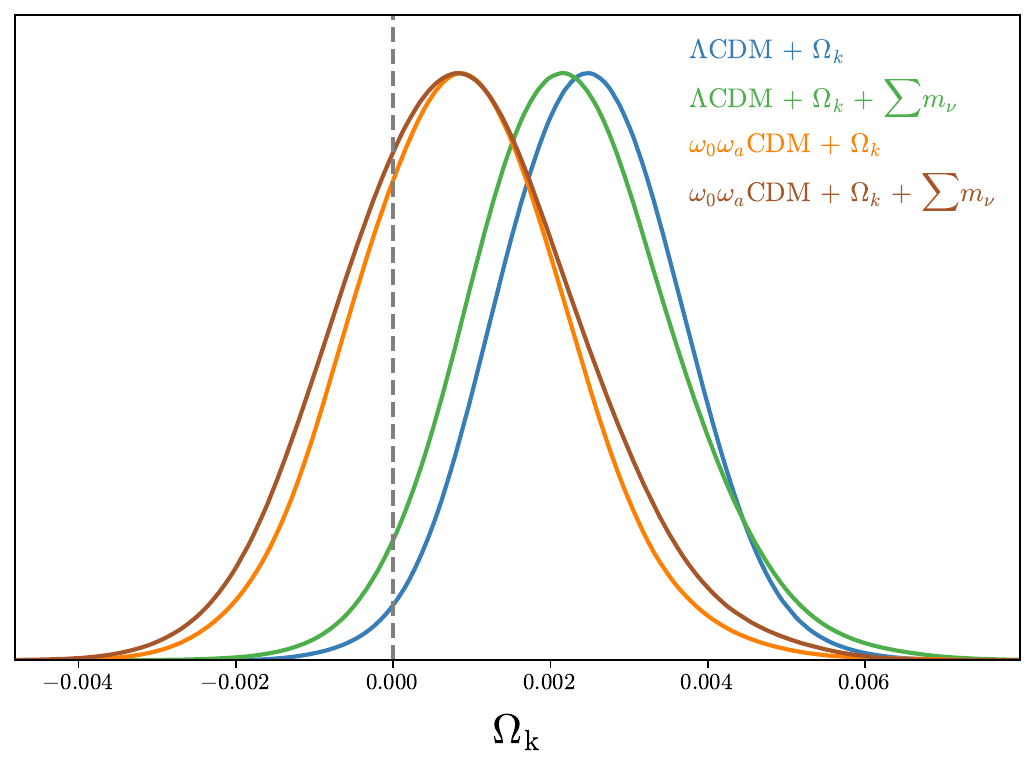}
     \end{subfigure}
     \caption{(\textbf{Upper}) 1D posterior distribution for $\sum m_\nu$ across different models. (\textbf{Lower}) 1D posterior distribution for $\Omega_k$ across different models. The vertical dashed lines indicate $\sum m_\nu =0.06$ eV and $\Omega_k=0$, respectively.}
     \label{fig:1D_Pos_Posteriors}
\end{figure}  

\newpage

The left panel of Figure \ref{fig:1D_Pos_Posteriors} shows that these models are 
unable to resolve the issue of the $\sum m_\nu$ posterior peak being located at 
positive values; instead, the distributions suggest a trend toward negative 
values. Since these posteriors are non-Gaussian, it is not possible to determine 
whether the shift observed when including curvature or switching models is 
physically motivated or a statistical artifact arising from the 
additional parameters. To explore this, we widen the 
prior to include the negative mass region. We performed additional runs following 
the methodology described above. Table \ref{tab:Results_reduced} shows our 
best-fit results for the main cosmological parameters, whereas in Appendix \ref{tab:Results_all} we display
the complete results for all sampled parameters across the analyzed models.

\begin{table*}[!t]
\centering
\begin{ruledtabular}
\makebox[\textwidth][c]{%
    \resizebox{1\textwidth}{!}{%
\begin{tabular}{l c c c c c c c c c c c c}
\toprule
\textbf{Model} & $\mathbf{\Omega_m}$ & $\mathbf{10^3 \Omega_k}$ & $\mathbf{\sum m_\nu}$ [eV] & $\mathbf{\omega_0}$ & $\mathbf{\omega_a}$ \\
\midrule
$\mathbf{\Lambda}$CDM  & $0.3049\pm 0.0036$ & - & - & - & -  \\
$\mathbf{\Lambda}$CDM + $\mathbf{\sum m_\nu}$ & $0.3035\pm 0.0036$ & - & $< 0.0731$ & - & -  \\
$\mathbf{\Lambda}$CDM + $\mathbf{\Omega_k}$ & $0.3058\pm 0.0036$ & $2.5\pm 1.2$ & - & - & -  \\
$\mathbf{\Lambda}$CDM + $\mathbf{\Omega_k}$ + $\mathbf{\sum m_\nu}$ & $0.3055^{+0.0039}_{-0.0044}$ & $2.3^{+1.2}_{-1.4}$ & $< 0.135$ & - & -  \\
$\mathbf{\omega_0\omega_a}$CDM + $\mathbf{\sum m_{\nu}}$ & $0.3187\pm 0.0057$ & - & $< 0.130$ & -$0.758\pm 0.059$ & -$0.82^{+0.26}_{-0.21}$  \\
$\mathbf{\omega_0\omega_a}$CDM + $\mathbf{\Omega_k}$ & $0.3187\pm 0.0059$ & $0.9\pm 1.4$ & - & -$0.774^{+0.073}_{-0.057}$ & -$0.76^{+0.23}_{-0.29}$  \\
$\mathbf{\omega_0\omega_a}$CDM + $\mathbf{\Omega_k}$ + $\mathbf{\sum m_\nu}$ & $0.3190\pm 0.0061$ & $0.9^{+1.4}_{-1.6}$ & $< 0.171$ & -$0.773^{+0.074}_{-0.060}$ & -$0.76^{+0.26}_{-0.30}$  \\
\midrule
$\mathbf{\Lambda}$CDM + $\mathbf{\sum m_{\nu,eff}}$ & $0.2992\pm 0.0045$ & - & -$0.073^{+0.051}_{-0.064}$ & - & -  \\
$\mathbf{\Lambda}$CDM + $\mathbf{\Omega_k}$ + $\mathbf{\sum m_{\nu,eff}}$ & $0.3013\pm 0.0054$ & $1.1^{+1.5}_{-1.7}$ & -$0.011^{+0.052}_{-0.050}$ & - & -  \\
$\mathbf{\omega_0\omega_a}$CDM + $\mathbf{\sum m_{\nu,eff}}$ & $0.3156\pm 0.0063$ & - & -$0.052^{+0.095}_{-0.084}$ & -$0.787\pm 0.061$ & -$0.63^{+0.29}_{-0.26}$  \\
$\mathbf{\omega_0\omega_a}$CDM + $\mathbf{\Omega_k}$ + $\mathbf{\sum m_{\nu,eff}}$ & $0.3146\pm 0.0071$ & -$0.6^{+1.6}_{-1.9}$ & -$0.07\pm 0.11$ & -$0.786\pm 0.061$ & -$0.61^{+0.29}_{-0.26}$  \\
\bottomrule
\end{tabular}%
}
}
\caption{Key cosmological parameter constraints. Results represent the 68\% C.L. intervals, except for the sum of neutrino masses, which is quoted at the 95\% C.L. Regarding the results for the $\mathbf{\omega_0\omega_a}$CDM + $\mathbf{\sum m_{\nu}}$ model, the analysis utilizes the publicly available MCMC chains provided by the DESI collaboration as part of their Data Release 2 (DR2) \cite{DESI_DR2_II}.}
\label{tab:Results_reduced}
\end{ruledtabular}
\end{table*}

\begin{subsection}{Effective neutrino mass results}\label{Sec:negnu_Results}

A particle with negative mass is physically meaningless. However, in our context, it serves as a means to explore a new parameter space which, as suggested in Figure \ref{fig:1D_Pos_Posteriors}, is statistically preferred by the data. Figure \ref{fig:matriz_triangles} presents triangular plots for the four analyzed models, comparing them with their counterparts that are subject to positive priors. It is clear that extending the parameter space to include negative masses preserves the inherent degeneracies of the neutrino mass sum, while the transition between these regimes remains smooth. Note that all posteriors that include negative mass priors peak at the negative mass range. This is clearly seen in Figure \ref{fig:Negnu_1D_Mnu} that summarizes the 1D probability behavior of $\sum m_{\nu,eff}$ across all cases. The white region, corresponding to a positive neutrino mass, yields results equivalent to those shown in the left panel of Figure \ref{fig:1D_Pos_Posteriors}. However, as previously noted, the most interesting and complementary insights emerge in the region where $\sum m_\nu <0$. The posterior for the $\omega_0\omega_a$CDM+$\Omega_k$+$\sum m_{\nu,eff}$ model exhibits a larger standard deviation, and hence, yields a larger positive mass area; nevertheless, the peak of this distribution is shifted further away from the lower bound of the neutrino mass sum. The $\Lambda$CDM + $\sum m_{\nu,eff}$ model appears to exhibit the poorest performance, showing a $2.59\sigma$ tension relative to the value $\sum m_\nu =0.06$ eV. Conversely, the $\Lambda$CDM + $\Omega_k$ + $\sum m_{\nu,eff}$ model shows a better performance, yielding a tension of $1.17\sigma$ with respect to the same value. Regarding the $\omega_0\omega_a$CDM models, $\omega_0\omega_a$CDM + $\sum m_{\nu,eff}$ shows a $1.18\sigma$ tension while $\omega_0\omega_a$CDM + $\Omega_k$ + $\sum m_{\nu,eff}$ can reduce it to $1.13\sigma$ losing constraining power in exchange.

As shown in Appendix \ref{Apendix: Dovekie_Results}, the constraints on $\sum m_\nu$ are insensitive to the updated reanalysis for the DESY5 dataset. This reinforces the overall robustness of the findings presented above.

\begin{figure}[!t]
    \centering
    \includegraphics[width=\linewidth]{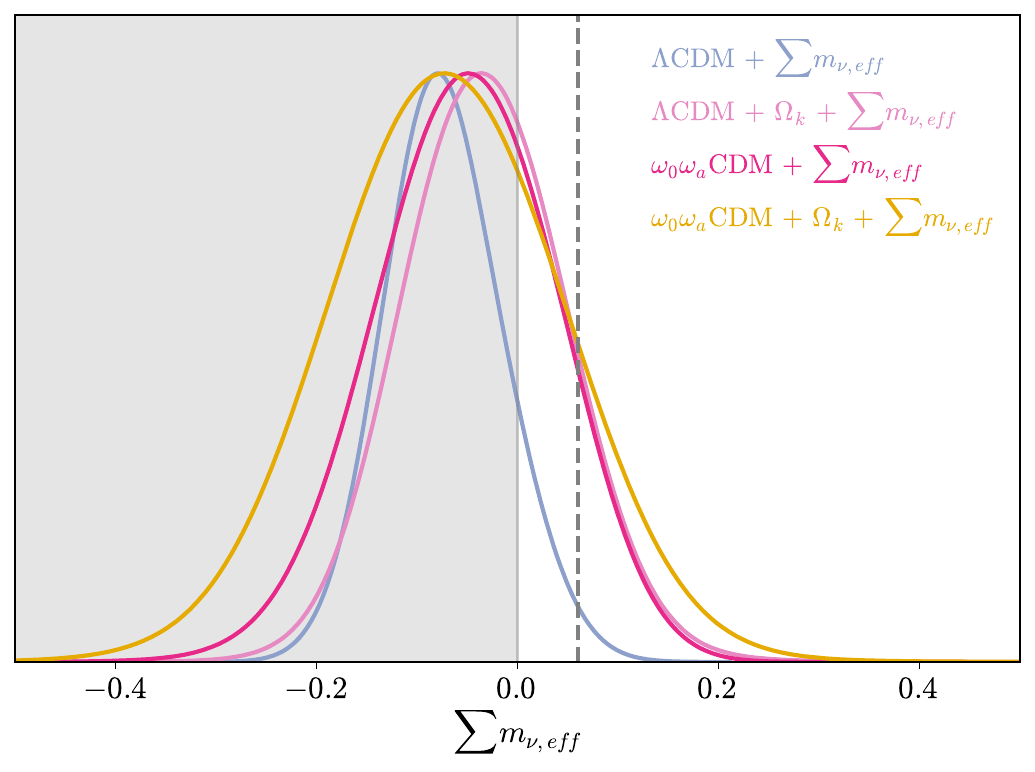}
    \caption{1D posterior distributions for the effective neutrino mass sum, $\sum m_{\nu,eff}$. Interestingly, while the behavior in the positive region remains consistent with results obtained under a strictly positive prior, the distribution peaks in the negative mass range indicate that the $\Lambda$CDM+$\Omega_k$+$\sum m_{\nu,eff}$ model exhibits the best performance. Once again, we observe that the neutrino mass constraints for the $\Lambda$CDM+$\Omega_k$+$\sum m_{\nu,eff}$ and $\omega_0\omega_a$CDM+$\sum m_{\nu,eff}$ models are remarkably similar. The vertical dashed line denotes the value of $\sum m_\nu = 0.06$ eV, and the grey shaded area highlights the negative mass region.}
    \label{fig:Negnu_1D_Mnu}
\end{figure}

\end{subsection}

\begin{subsection}{Goodness of fit}\label{Sec:Fit_Results}

To evaluate model performance beyond parameter estimation alone and to determine the goodness-of-fit to the data, we computed several statistical metrics based on the results obtained from the \texttt{iminuit} optimizer. Specifically, we report the $\Delta\chi^2_{MAP}$ and the Akaike Information Criterion (AIC), both based on the maximum likelihood; additionally, we provide the Deviance Information Criterion (DIC) derived from the posterior samples, following the standard methodology described in Ref. \cite{Liddle_2007}.

First, the fit is assessed via the Maximum a Posteriori (MAP) estimate, quantified by the log-likelihood difference $\Delta\chi^2_{MAP} \equiv \text{-}2\Delta\ln{\mathcal{L}}$. To compare the relative performance of models with varying complexities, we employ the AIC, defined as:
\begin{equation}
    AIC \equiv \chi^2 + 2k \, , 
\end{equation}
where $k$ denotes the number of free parameters.

Additionally, the DIC is utilized to leverage the full posterior information provided by the MCMC chains. A key step in this calculation is defining the effective number of model parameters, $p_D$. In this work, we specifically adopt the alternative form for this quantity by setting $\chi^2(\overline{\theta})=\chi^2_{min}(\theta)$, which yields:
\begin{equation}
    p_{D} \equiv \overline{\chi^2(\theta)}\text{ - }\chi^2_{min}(\theta) \, ,
\end{equation}
that leads to the expression:
\begin{equation}\label{Eq:DIC}
    DIC \equiv 2\overline{\chi^2(\theta)}\text{ - }\chi^2_{min}(\theta) \,.
\end{equation}

\bigskip

Finally, we compute the \textit{Bayesian Evidence} and derive the corresponding Bayes factor as an additional criterion for model selection. The Bayesian evidence is defined as \cite{Trotta_2008}:

\begin{equation}
    \mathcal{Z} = \int p\left( d|\theta \right)p\left( \theta \right)d\theta
\end{equation}
where $p\left( d|\theta \right)$ is the likelihood function and $p\left( \theta \right)$ is the prior. For model selection we compute the Bayes factor, defined as
\begin{equation}
    \mathcal{B}_{0,1} = \ln{\frac{\mathcal{Z}_0}{\mathcal{Z}_1}}
\end{equation}
A positive Bayes factor value indicates a preference for the model 0.
The criteria are: If $\mathcal{B}_{0,1}<1$, the evidence is not conclusive; however, if $\mathcal{B}_{0,1}>5$, there is strong evidence favoring model 0 over model 1. See Table 1 of \cite{Trotta_2008} for more details.

Computing the Bayesian evidence is challenging due to the high dimensionality of the parameter space, particularly when using unmarginalized CMB likelihoods. To address this issue, we employed the \texttt{harmonic} code \cite{mcewen2023machinelearningassistedbayesian}.

Table \ref{tab:GoF} summarizes the goodness-of-fit metrics using $\Lambda$CDM as the reference model. For the information criteria, we computed $\Delta X = X_{model}\text{-}X_{\Lambda CDM}$, where $X$ is the chosen criterion ($\chi^2_{MAP}$, AIC or DIC). For these metrics, a lower value indicates a superior model fit. Conversely, the Bayes factor is defined as $\mathcal{B}_{0,1} = \ln{\mathcal{Z}_{\Lambda CDM}- \ln\mathcal{Z}_{model}}$. Thus, a positive $\mathcal{B}_{0,1}$ value indicates a preference for the reference $\Lambda$CDM model.

Regarding the $\Delta\chi^2_{MAP}$ values, the $\omega_0\omega_a$CDM model and its extensions yield improved fits, which is expected due to the additional degrees of freedom. Furthermore, the $\Delta$AIC results indicate a prominent improvement for the $\omega_0\omega_a$CDM models with fixed neutrino mass relative to $\Lambda$CDM, while the $\Delta$DIC results, though less pronounced, consistently favor the $\omega_0\omega_a$CDM framework.

Interestingly, the Bayes factor points in the opposite direction, strongly favoring the minimal $\Lambda$CDM model. On the other hand, a general improvement of the fit in the negative neutrino-mass region is observed in both $\Delta\chi^2_{\rm MAP}$ and $\Delta$AIC. Allowing this expanded portion of the parameter space accommodates the general tendency of the datasets to favor this region, yielding smaller $\chi^2$ values than those obtained in the strictly positive-mass cases, with the exception of the $\Lambda$CDM model featuring spatial curvature and free neutrino mass. Nevertheless, despite these improved maximum likelihood estimates, the Bayesian evidence heavily penalizes this expanded parameter volume, maintaining a robust statistical preference against the negative-mass region.

\begin{table}[!t]
\centering
\resizebox{0.49\textwidth}{!}{%
\begin{tabular}{l c c c c}
    \toprule
    \textbf{Model} & $\mathbf{\Delta\chi^2_{\text{MAP}}}$ & $\mathbf{\Delta}$AIC & $\mathbf{\Delta}$DIC & $\mathbf{\mathcal{B}_{0,1}}$ \\
    \midrule
    $\Lambda$CDM + $\Omega_k$ & -10.48 & -8.48 & -3.01 & 3.10 \\
    $\Lambda$CDM + $\sum m_{\nu}$ & -2.81 & -0.81 & 4.03 & 3.17 \\
    $\Lambda$CDM + $\sum m_{\nu,eff}$ & -6.21 & -4.21 & 1.91 & 7.69 \\
    $\Lambda$CDM + $\Omega_k + \sum m_{\nu}$ & -8.79 & -4.79 & 1.64 & 6.72 \\
    $\Lambda$CDM + $\Omega_k + \sum m_{\nu,eff}$ & -7.3 & -3.30 & 1.47 & 7.57 \\
    
    \midrule 
    
    $\omega_0\omega_a$CDM & -6.35 & -2.35 & - & - \\
    $\omega_0\omega_a$CDM + $\Omega_k$ & -21.06 & -15.06 & -6.69 & 1.33 \\
    $\omega_0\omega_a$CDM + $\sum m_{\nu}$ & -10.09 & -4.09 & -4.01 & 0.11 \\
    $\omega_0\omega_a$CDM + $\sum m_{\nu,eff}$ & -15.87 & -9.87 & -5.68 & 9.70 \\
    $\omega_0\omega_a$CDM + $\Omega_k + \sum m_{\nu}$ & -16.26 & -6.26 & -8.26 & 4.94 \\
    $\omega_0\omega_a$CDM + $\Omega_k + \sum m_{\nu,eff}$ & -18.64 & -10.64 & -3.83 & 5.37 \\
    
    \bottomrule
\end{tabular}
}
\caption{Comparison of statistical criteria relative to the $\Lambda$CDM model as reference. Differences are calculated as $\Delta X = X_{model}\text{-}X_{\Lambda CDM}$. For the Bayes factor, $\mathcal{B}_{0,1} = \ln \mathcal{Z}_{\Lambda CDM} \text{-} \ln \mathcal{Z}_{model}$}
\label{tab:GoF}
\end{table}

For any given cosmology, models restricted to positive neutrino masses are explicitly preferred by the Bayesian evidence over their counterparts allowing for negative masses. This demonstrates that the negative mass region does not provide a statistically superior global representation of the data.

Although negative neutrino masses are not preferred by the metrics, our results show that the posterior distribution peaks at negative (unphysical) values in all the cases studied in this work.

\end{subsection}

\end{section}

\begin{section}{Conclusions}\label{Sec:Conclusions}

The $\Lambda$CDM model remains under constant scrutiny in the current 
era of precision cosmology, facing significant challenges such as the one
addressed here, i.e. the apparent discrepancy between cosmological upper bounds 
on the neutrino mass sum and the lower limits established by terrestrial 
oscillation experiments. The primary objective of this work was to analyze 
the impact of spatial curvature on neutrino mass constraints by exploring 
the degeneracies within the parameter space. To this end, we employed the 
latest CMB, BAO, and SNe Ia datasets to perform Bayesian parameter 
estimation for the $\Lambda$CDM and $\omega_0\omega_a$CDM models, alongside 
their respective extensions including $\Omega_k$ and $\sum m_\nu$. This analysis 
included an exploration of the parameter space where the neutrino 
mass takes negative effective values, allowing for a deeper evaluation 
of the trends observed in the data.

In the scenario where a strictly positive neutrino mass was imposed, our 
results are consistent with previous literature. The inclusion of spatial 
curvature as a free parameter shows an apparent reduction in the tension 
with terrestrial measurements for the $\Lambda$CDM model. This effect is 
comparable to that observed when shifting toward a dynamical dark energy 
equation of state, as illustrated in the left panel of Figure 
\ref{fig:1D_Pos_Posteriors}. However, since the resulting posteriors 
in these models exhibit a strong truncation at $\sum m_\nu = 0$ and a 
marked departure from Gaussianity, it is not possible to quantify this 
tension through the conventional $\sigma$ standard deviation 
method. Consequently, one cannot determine whether the observed 
reduction in tension represents a physical convergence toward the 
experimental lower limit, $0.06$ eV, or is merely a consequence 
of the loss in constraining power as the model complexity 
increases. This statistical ambiguity is precisely what motivates 
the shift toward the negative effective mass scenario. 

\begin{figure*}[!t]
    \centering
    \begin{subfigure}[b]{0.49\textwidth}
        \centering
        \includegraphics[width=\textwidth]{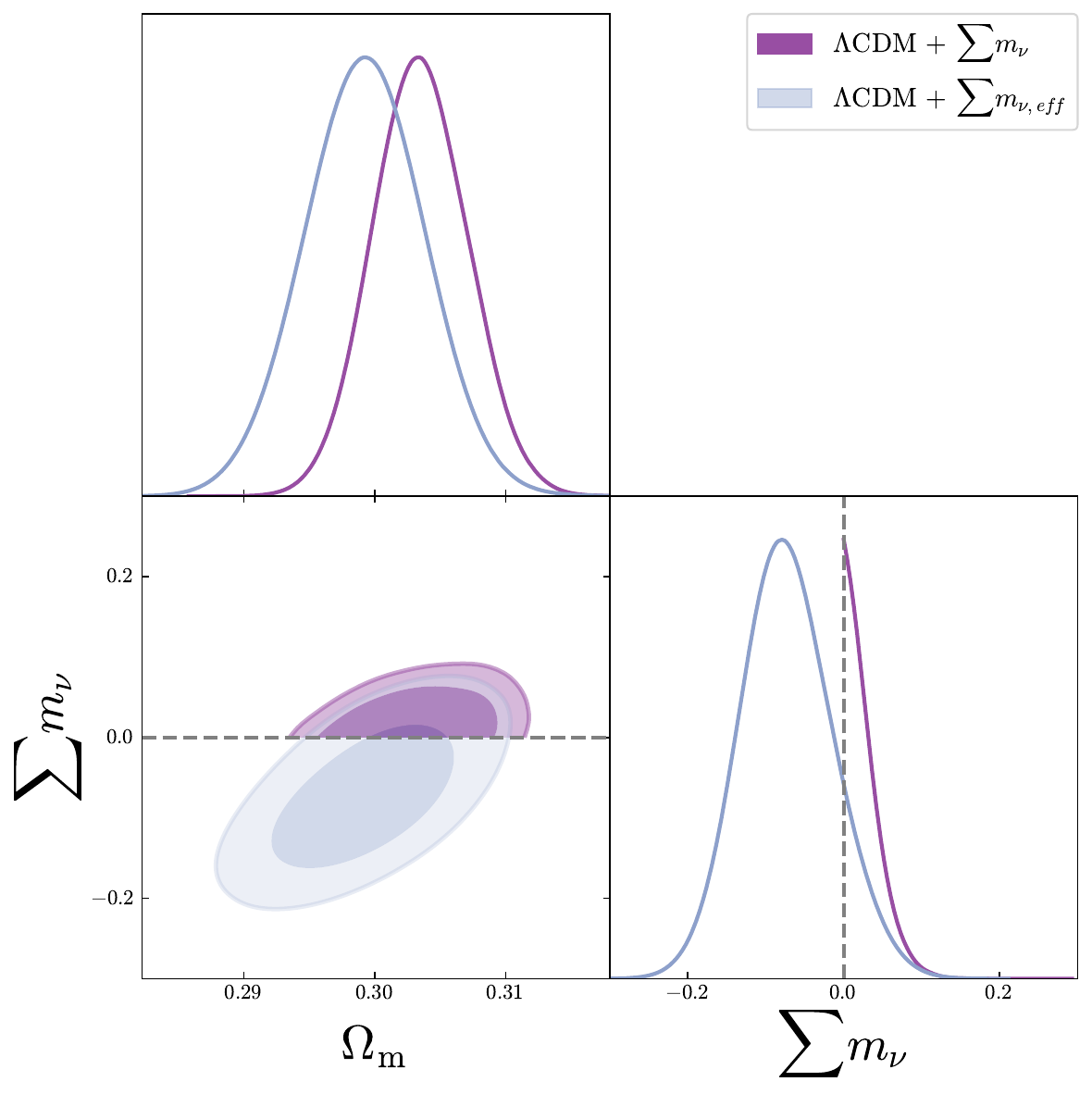}
    \end{subfigure}
    \hfill 
    \begin{subfigure}[b]{0.49\textwidth}
        \centering
        \includegraphics[width=\textwidth]{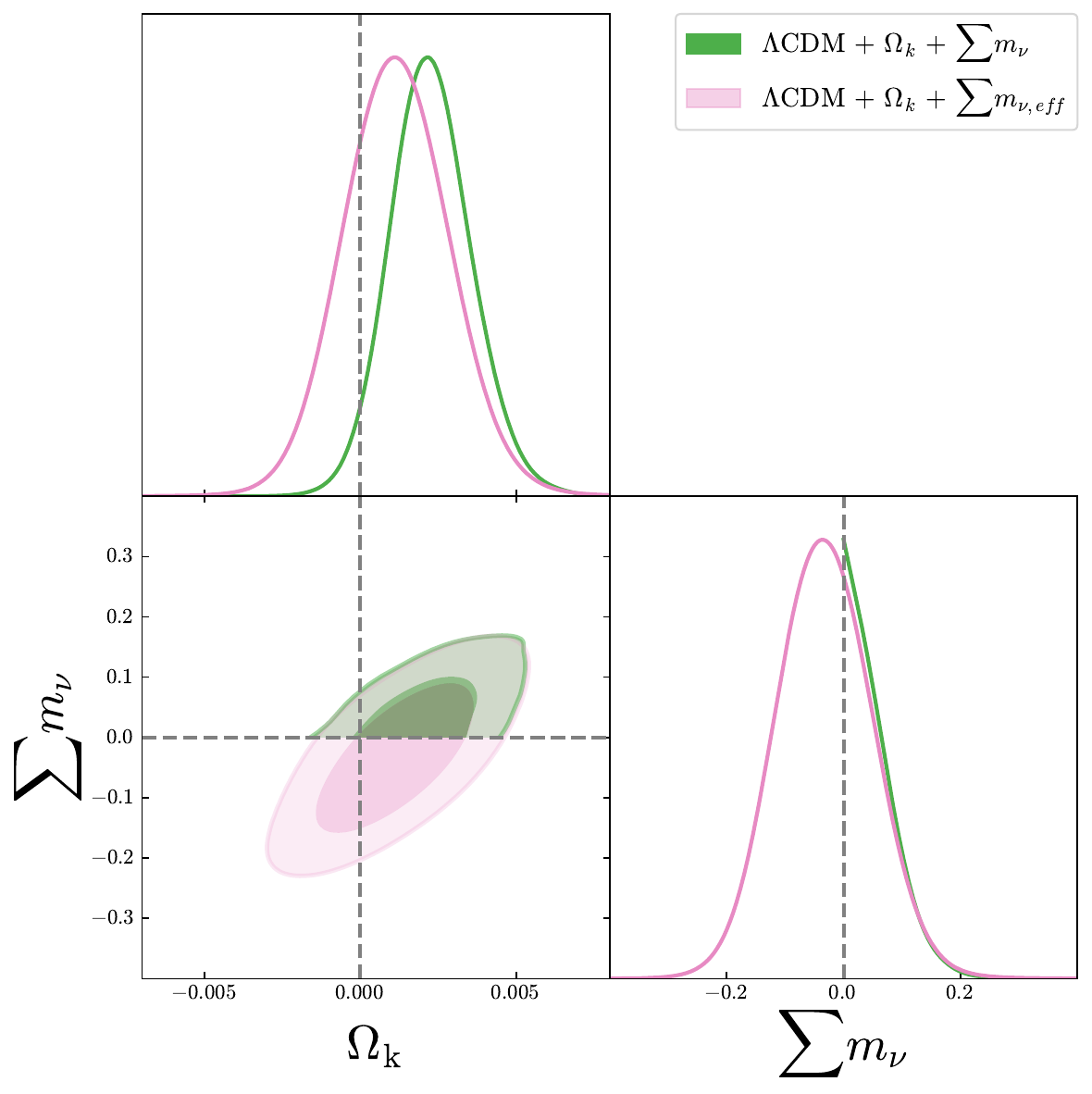}
    \end{subfigure}

    \vspace{0.5cm} 

    \begin{subfigure}[b]{0.49\textwidth}
        \centering
        \includegraphics[width=\textwidth]{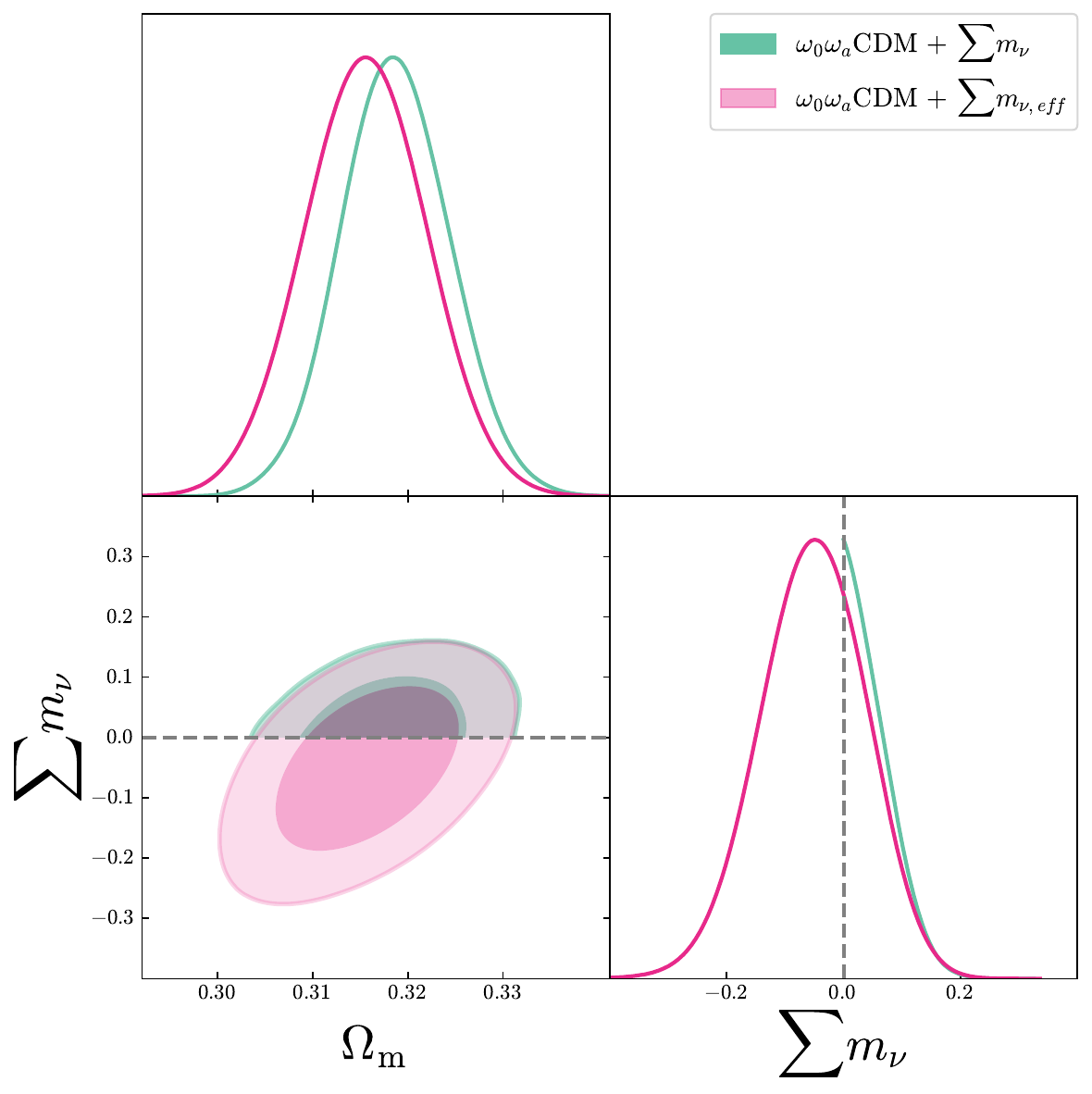}
    \end{subfigure}
    \hfill
    \begin{subfigure}[b]{0.49\textwidth}
        \centering
        \includegraphics[width=\textwidth]{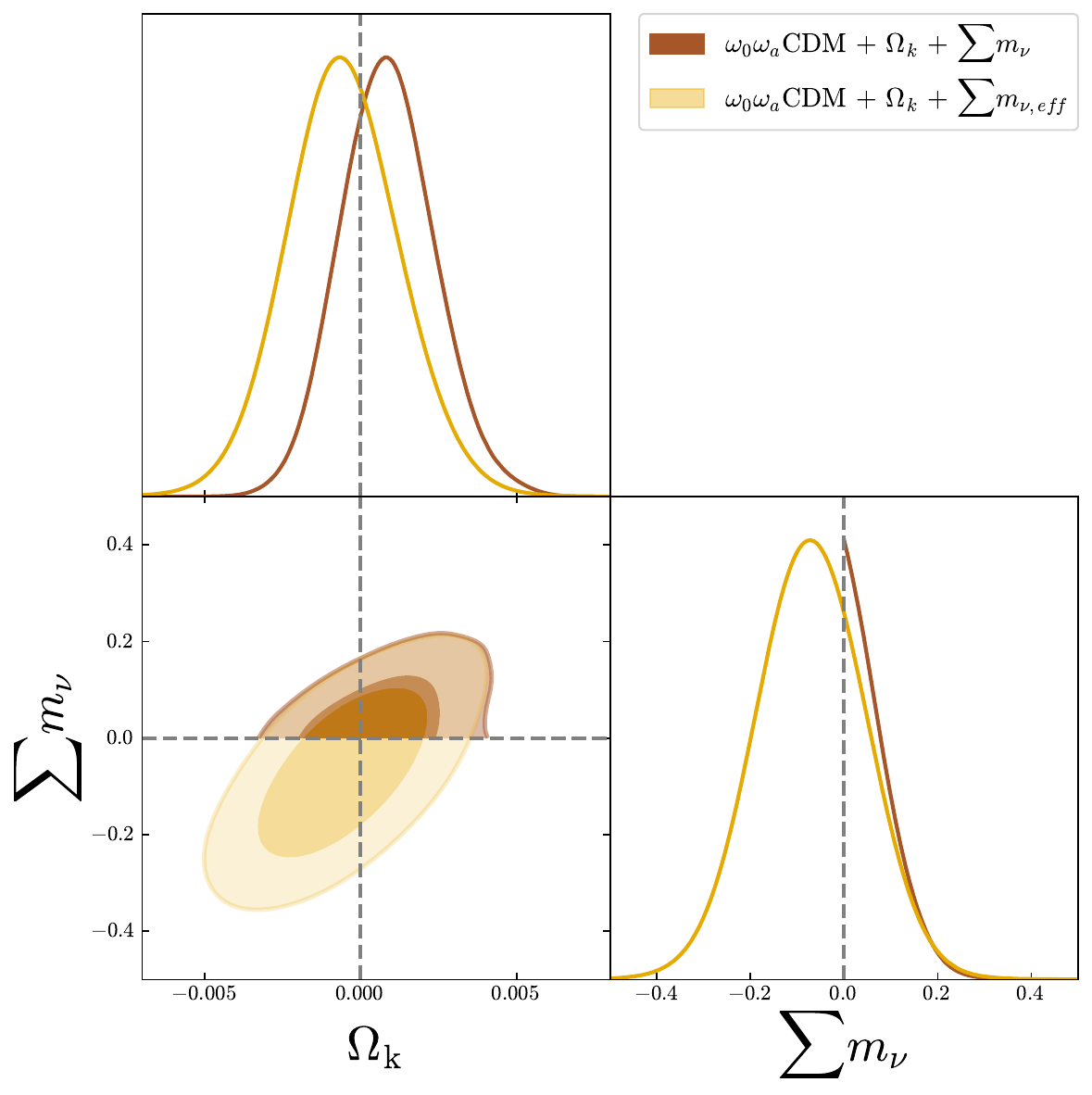}
    \end{subfigure}

    \caption{Illustrative posterior distribution for the four analyzed scenarios. The smooth behavior of the contours across the different panels reflects the preference for the negative mass regime and the robustness of the sampling process. Notably, the characteristic degeneracies between cosmological parameters are consistently maintained throughout the various models extensions, confirming that the physical correlations remain stable even as the parameter space is expanded.} 
    \label{fig:matriz_triangles}
\end{figure*}

\clearpage

To address this issue, we modified the communication interface between the Boltzmann code (CLASS) and the sampler (Cobaya), enabling an agnostic exploration of the mass sign. A key result presented in Figure \ref{fig:matriz_triangles} is the existence of a smooth transition in the posteriors between the positive and negative mass regimes. This continuity underscores that allowing $\sum m_\nu < 0$ provides a useful diagnostic of boundary-induced effects for these datasets; otherwise, the parameter space is artificially constrained, introducing biases in the estimates. From our results, we observe that curvature reduces the tension relative to the $0.06$ eV value, bringing it down from $2.59\sigma$, given $\sum m_{\nu,eff} = -0.073^{+0.051}_{-0.064}$ in the $\Lambda$CDM + $\sum m_{\nu,eff}$ model, to $1.17\sigma$ with $\sum m_{\nu,eff} = -0.011^{+0.052}_{-0.050}$ within the $\Lambda$CDM + $\Omega_k$ + $\sum m_{\nu,eff}$ model. While this trend persists in the $\omega_0\omega_a$CDM + $\sum m_{\nu,eff}$ model, the increased number of free parameters degrades the overall constraining power. Analysis of the various goodness-of-fit criteria reveals an interesting dichotomy. On the one hand, metrics based on maximum likelihood and posterior-based criteria (such as AIC and DIC) suggest that the $\omega_0\omega_a$CDM model and its extensions provide a better fit due to their greater parameter flexibility, although this does not inherently resolve the preference for null or negative mass values. On the other hand, the Bayesian evidence strongly opposes this trend. The Bayes factor strongly penalizes more complex models and therefore consistently favors the baseline $\Lambda$CDM models over their dynamical dark energy counterparts, and explicitly penalizes the extensions that allow for negative masses. Ultimately, while models like $\Lambda$CDM + $\Omega_k$ + $\sum m_{\nu}$ and $\omega_0\omega_a$CDM + $\sum m_\nu$ perform similarly, the statistical advantage provided by expanded parameter volumes allows the negative mass region to yield a noticeably better fit to the data. However, the rigorous Bayesian penalty for this extended volume prevents it from being statistically recognized as a superior global model. Nevertheless, the posterior distribution of the neutrino mass exhibits a maximum at negative (unphysical) values in all cases studied in this work.

The recent trend of questioning the validity of the standard model, largely driven by the DESI results, has initiated a period of intense activity in modern cosmology. Our findings suggest that the use of negative effective masses should not be viewed as a mere mathematical curiosity but should be considered a diagnostic tool in cosmological parameter analysis. This practice allows for the identification of prior-driven biases and ensures that conclusions regarding the nature of neutrinos are not artifacts of an arbitrary delimitation of the parameter space. In a landscape where tensions between CMB, BAO, and SNe Ia datasets suggest possible departures from $\Lambda$CDM, adopting methodologies that are agnostic to physical boundaries is important for assessing whether we are witnessing new physics or statistical fluctuations at the limit of our current precision.

\end{section}

\begin{section}{Acknowledgments}
The authors acknowledge support by SECIHTI (CONAHCyT) project CBF2023-2024-589. H. P-H acknowledges support from SECIHTI (CONAHCyT) through the doctoral scholarship, CVU No. 1182916.

This research used data obtained with the Dark Energy Spectroscopic Instrument (DESI). DESI construction and operations is managed by the Lawrence Berkeley National Laboratory. This material is based upon work supported by the \href{https://www.energy.gov/}{U.S. Department of Energy}, Office of Science, Office of High-Energy Physics, under Contract No. DE–AC02–05CH11231, and by the National Energy Research Scientific Computing Center, a DOE Office of Science User Facility under the same contract. Additional support for DESI was provided by the \href{https://www.nsf.gov/}{U.S. National Science Foundation} (NSF), Division of Astronomical Sciences under Contract No. AST-0950945 to the NSF’s National Optical-Infrared Astronomy Research Laboratory; the \href{https://www.ukri.org/councils/stfc/}{Science and Technology Facilities Council of the United Kingdom}; the \href{https://www.moore.org/}{Gordon and Betty Moore Foundation}; the \href{https://www.hsfoundation.org/}{Heising-Simons Foundation}; the \href{https://www.cea.fr/}{French Alternative Energies and Atomic Energy Commission} (CEA); the \href{https://secihti.mx/}{National Council of Humanities, Science and Technology of Mexico} (CONAHCYT); the \href{https://portal.mineco.gob.es/es-es/Paginas/index.aspx}{Ministry of Science and Innovation of Spain} (MICINN), and by the DESI Member Institutions: www.desi.lbl.gov/collaborating-institutions. The DESI collaboration is honored to be permitted to conduct scientific research on I’oligam Du’ag (Kitt Peak), a mountain with particular significance to the \href{https://www.tonation-nsn.gov/}{Tohono O’odham Nation}. Any opinions, findings, and conclusions or recommendations expressed in this material are those of the author(s) and do not necessarily reflect the views of the U.S. National Science Foundation, the U.S. Department of Energy, or any of the listed funding agencies.

\end{section}

\newpage
 \bibliographystyle{apsrev}
 \bibliography{neutrino}

@article{Planck2018_Results,
   title={Planck2018 results: V. {CMB} power spectra and likelihoods},
   volume={641},
   ISSN={1432-0746},
   url={http://dx.doi.org/10.1051/0004-6361/201936386},
   DOI={10.1051/0004-6361/201936386},
   journal={Astronomy \& Astrophysics},
   publisher={EDP Sciences},
   author={Aghanim, N. and Akrami, Y. and Ashdown, M. and Aumont, J. and Baccigalupi, C. and Ballardini, M. and Banday, A. J. and Barreiro, R. B. and Bartolo, N. and Basak, S. and Benabed, K. and Bernard, J.-P. and Bersanelli, M. and Bielewicz, P. and Bock, J. J. and Bond, J. R. and Borrill, J. and Bouchet, F. R. and Boulanger, F. and Bucher, M. and Burigana, C. and Butler, R. C. and Calabrese, E. and Cardoso, J.-F. and Carron, J. and Casaponsa, B. and Challinor, A. and Chiang, H. C. and Colombo, L. P. L. and Combet, C. and Crill, B. P. and Cuttaia, F. and de Bernardis, P. and de Rosa, A. and de Zotti, G. and Delabrouille, J. and Delouis, J.-M. and Di Valentino, E. and Diego, J. M. and Doré, O. and Douspis, M. and Ducout, A. and Dupac, X. and Dusini, S. and Efstathiou, G. and Elsner, F. and Enßlin, T. A. and Eriksen, H. K. and Fantaye, Y. and Fernandez-Cobos, R. and Finelli, F. and Frailis, M. and Fraisse, A. A. and Franceschi, E. and Frolov, A. and Galeotta, S. and Galli, S. and Ganga, K. and Génova-Santos, R. T. and Gerbino, M. and Ghosh, T. and Giraud-Héraud, Y. and González-Nuevo, J. and Górski, K. M. and Gratton, S. and Gruppuso, A. and Gudmundsson, J. E. and Hamann, J. and Handley, W. and Hansen, F. K. and Herranz, D. and Hivon, E. and Huang, Z. and Jaffe, A. H. and Jones, W. C. and Keihänen, E. and Keskitalo, R. and Kiiveri, K. and Kim, J. and Kisner, T. S. and Krachmalnicoff, N. and Kunz, M. and Kurki-Suonio, H. and Lagache, G. and Lamarre, J.-M. and Lasenby, A. and Lattanzi, M. and Lawrence, C. R. and Le Jeune, M. and Levrier, F. and Lewis, A. and Liguori, M. and Lilje, P. B. and Lilley, M. and Lindholm, V. and López-Caniego, M. and Lubin, P. M. and Ma, Y.-Z. and Macías-Pérez, J. F. and Maggio, G. and Maino, D. and Mandolesi, N. and Mangilli, A. and Marcos-Caballero, A. and Maris, M. and Martin, P. G. and Martínez-González, E. and Matarrese, S. and Mauri, N. and McEwen, J. D. and Meinhold, P. R. and Melchiorri, A. and Mennella, A. and Migliaccio, M. and Millea, M. and Miville-Deschênes, M.-A. and Molinari, D. and Moneti, A. and Montier, L. and Morgante, G. and Moss, A. and Natoli, P. and Nørgaard-Nielsen, H. U. and Pagano, L. and Paoletti, D. and Partridge, B. and Patanchon, G. and Peiris, H. V. and Perrotta, F. and Pettorino, V. and Piacentini, F. and Polenta, G. and Puget, J.-L. and Rachen, J. P. and Reinecke, M. and Remazeilles, M. and Renzi, A. and Rocha, G. and Rosset, C. and Roudier, G. and Rubiño-Martín, J. A. and Ruiz-Granados, B. and Salvati, L. and Sandri, M. and Savelainen, M. and Scott, D. and Shellard, E. P. S. and Sirignano, C. and Sirri, G. and Spencer, L. D. and Sunyaev, R. and Suur-Uski, A.-S. and Tauber, J. A. and Tavagnacco, D. and Tenti, M. and Toffolatti, L. and Tomasi, M. and Trombetti, T. and Valiviita, J. and Van Tent, B. and Vielva, P. and Villa, F. and Vittorio, N. and Wandelt, B. D. and Wehus, I. K. and Zacchei, A. and Zonca, A.},
   year={2020},
   month=sep, pages={A5} }

@article{CamSpec_1,
   title={{CMB} power spectra and cosmological parameters from Planck {PR4} with {CamSpec}},
   volume={517},
   ISSN={1365-2966},
   url={http://dx.doi.org/10.1093/mnras/stac2744},
   DOI={10.1093/mnras/stac2744},
   number={3},
   journal={Monthly Notices of the Royal Astronomical Society},
   publisher={Oxford University Press (OUP)},
   author={Rosenberg, Erik and Gratton, Steven and Efstathiou, George},
   year={2022},
   month=sep, pages={4620––4636} }

@article{Tristram_2021,
   title={Planck constraints on the tensor-to-scalar ratio},
   volume={647},
   ISSN={1432-0746},
   url={http://dx.doi.org/10.1051/0004-6361/202039585},
   DOI={10.1051/0004-6361/202039585},
   journal={Astronomy \& Astrophysics},
   publisher={EDP Sciences},
   author={Tristram, M. and Banday, A. J. and Górski, K. M. and Keskitalo, R. and Lawrence, C. R. and Andersen, K. J. and Barreiro, R. B. and Borrill, J. and Eriksen, H. K. and Fernandez-Cobos, R. and Kisner, T. S. and Martínez-González, E. and Partridge, B. and Scott, D. and Svalheim, T. L. and Thommesen, H. and Wehus, I. K.},
   year={2021},
   month=mar, pages={A128} }

@article{Tristram_2024_PR4,
   title={Cosmological parameters derived from the final {Planck} data release ({PR4})},
   volume={682},
   ISSN={1432-0746},
   url={http://dx.doi.org/10.1051/0004-6361/202348015},
   DOI={10.1051/0004-6361/202348015},
   journal={Astronomy \& Astrophysics},
   publisher={EDP Sciences},
   author={Tristram, M. and Banday, A. J. and Douspis, M. and Garrido, X. and Górski, K. M. and Henrot-Versillé, S. and Hergt, L. T. and Ilić, S. and Keskitalo, R. and Lagache, G. and Lawrence, C. R. and Partridge, B. and Scott, D.},
   year={2024},
   month=jan, pages={A37} }

@article{Efstathiou_2021_CamSpecPipeline,
   title={A Detailed Description of the {CamSpec} Likelihood Pipeline and a Reanalysis of the {Planck} High Frequency Maps},
   volume={4},
   ISSN={2565-6120},
   url={http://dx.doi.org/10.21105/astro.1910.00483},
   DOI={10.21105/astro.1910.00483},
   number={1},
   journal={The Open Journal of Astrophysics},
   publisher={Maynooth University},
   author={Efstathiou, George and Gratton, Steven},
   year={2021},
   month=aug }

@article{Madhavacheril_2024_ACTDR6,
   title={The {Atacama} {Cosmology} {Telescope}: {DR6} Gravitational Lensing Map and Cosmological Parameters},
   volume={962},
   ISSN={1538-4357},
   url={http://dx.doi.org/10.3847/1538-4357/acff5f},
   DOI={10.3847/1538-4357/acff5f},
   number={2},
   journal={The Astrophysical Journal},
   publisher={American Astronomical Society},
   author={Madhavacheril, Mathew S. and Qu, Frank J. and Sherwin, Blake D. and MacCrann, Niall and Li, Yaqiong and Abril-Cabezas, Irene and Ade, Peter A. R. and Aiola, Simone and Alford, Tommy and Amiri, Mandana and Amodeo, Stefania and An, Rui and Atkins, Zachary and Austermann, Jason E. and Battaglia, Nicholas and Battistelli, Elia Stefano and Beall, James A. and Bean, Rachel and Beringue, Benjamin and Bhandarkar, Tanay and Biermann, Emily and Bolliet, Boris and Bond, J Richard and Cai, Hongbo and Calabrese, Erminia and Calafut, Victoria and Capalbo, Valentina and Carrero, Felipe and Challinor, Anthony and Chesmore, Grace E. and Cho, Hsiao-mei and Choi, Steve K. and Clark, Susan E. and Córdova Rosado, Rodrigo and Cothard, Nicholas F. and Coughlin, Kevin and Coulton, William and Crowley, Kevin T. and Dalal, Roohi and Darwish, Omar and Devlin, Mark J. and Dicker, Simon and Doze, Peter and Duell, Cody J. and Duff, Shannon M. and Duivenvoorden, Adriaan J. and Dunkley, Jo and Dünner, Rolando and Fanfani, Valentina and Fankhanel, Max and Farren, Gerrit and Ferraro, Simone and Freundt, Rodrigo and Fuzia, Brittany and Gallardo, Patricio A. and Garrido, Xavier and Givans, Jahmour and Gluscevic, Vera and Golec, Joseph E. and Guan, Yilun and Hall, Kirsten R. and Halpern, Mark and Han, Dongwon and Harrison, Ian and Hasselfield, Matthew and Healy, Erin and Henderson, Shawn and Hensley, Brandon and Hervías-Caimapo, Carlos and Hill, J. Colin and Hilton, Gene C. and Hilton, Matt and Hincks, Adam D. and Hložek, Renée and Ho, Shuay-Pwu Patty and Huber, Zachary B. and Hubmayr, Johannes and Huffenberger, Kevin M. and Hughes, John P. and Irwin, Kent and Isopi, Giovanni and Jense, Hidde T. and Keller, Ben and Kim, Joshua and Knowles, Kenda and Koopman, Brian J. and Kosowsky, Arthur and Kramer, Darby and Kusiak, Aleksandra and La Posta, Adrien and Lague, Alex and Lakey, Victoria and Lee, Eunseong and Li, Zack and Limon, Michele and Lokken, Martine and Louis, Thibaut and Lungu, Marius and MacInnis, Amanda and Maldonado, Diego and Maldonado, Felipe and Mallaby-Kay, Maya and Marques, Gabriela A. and McMahon, Jeff and Mehta, Yogesh and Menanteau, Felipe and Moodley, Kavilan and Morris, Thomas W. and Mroczkowski, Tony and Naess, Sigurd and Namikawa, Toshiya and Nati, Federico and Newburgh, Laura and Nicola, Andrina and Niemack, Michael D. and Nolta, Michael R. and Orlowski-Scherer, John and Page, Lyman A. and Pandey, Shivam and Partridge, Bruce and Prince, Heather and Puddu, Roberto and Radiconi, Federico and Robertson, Naomi and Rojas, Felipe and Sakuma, Tai and Salatino, Maria and Schaan, Emmanuel and Schmitt, Benjamin L. and Sehgal, Neelima and Shaikh, Shabbir and Sierra, Carlos and Sievers, Jon and Sifón, Cristóbal and Simon, Sara and Sonka, Rita and Spergel, David N. and Staggs, Suzanne T. and Storer, Emilie and Switzer, Eric R. and Tampier, Niklas and Thornton, Robert and Trac, Hy and Treu, Jesse and Tucker, Carole and Ullom, Joel and Vale, Leila R. and Van Engelen, Alexander and Van Lanen, Jeff and van Marrewijk, Joshiwa and Vargas, Cristian and Vavagiakis, Eve M. and Wagoner, Kasey and Wang, Yuhan and Wenzl, Lukas and Wollack, Edward J. and Xu, Zhilei and Zago, Fernando and Zheng, Kaiwen},
   year={2024},
   month=feb, pages={113} }

@article{ACT_DR6_Lensing,
   title={The {Atacama} {Cosmology} {Telescope}: A Measurement of the {DR6} {CMB} Lensing Power Spectrum and Its Implications for Structure Growth},
   volume={962},
   ISSN={1538-4357},
   url={http://dx.doi.org/10.3847/1538-4357/acfe06},
   DOI={10.3847/1538-4357/acfe06},
   number={2},
   journal={The Astrophysical Journal},
   publisher={American Astronomical Society},
   author={Qu, Frank J. and Sherwin, Blake D. and Madhavacheril, Mathew S. and Han, Dongwon and Crowley, Kevin T. and Abril-Cabezas, Irene and Ade, Peter A. R. and Aiola, Simone and Alford, Tommy and Amiri, Mandana and Amodeo, Stefania and An, Rui and Atkins, Zachary and Austermann, Jason E. and Battaglia, Nicholas and Battistelli, Elia Stefano and Beall, James A. and Bean, Rachel and Beringue, Benjamin and Bhandarkar, Tanay and Biermann, Emily and Bolliet, Boris and Bond, J Richard and Cai, Hongbo and Calabrese, Erminia and Calafut, Victoria and Capalbo, Valentina and Carrero, Felipe and Carron, Julien and Challinor, Anthony and Chesmore, Grace E. and Cho, Hsiao-mei and Choi, Steve K. and Clark, Susan E. and Córdova Rosado, Rodrigo and Cothard, Nicholas F. and Coughlin, Kevin and Coulton, William and Dalal, Roohi and Darwish, Omar and Devlin, Mark J. and Dicker, Simon and Doze, Peter and Duell, Cody J. and Duff, Shannon M. and Duivenvoorden, Adriaan J. and Dunkley, Jo and Dünner, Rolando and Fanfani, Valentina and Fankhanel, Max and Farren, Gerrit and Ferraro, Simone and Freundt, Rodrigo and Fuzia, Brittany and Gallardo, Patricio A. and Garrido, Xavier and Gluscevic, Vera and Golec, Joseph E. and Guan, Yilun and Halpern, Mark and Harrison, Ian and Hasselfield, Matthew and Healy, Erin and Henderson, Shawn and Hensley, Brandon and Hervías-Caimapo, Carlos and Hill, J. Colin and Hilton, Gene C. and Hilton, Matt and Hincks, Adam D. and Hložek, Renée and Ho, Shuay-Pwu Patty and Huber, Zachary B. and Hubmayr, Johannes and Huffenberger, Kevin M. and Hughes, John P. and Irwin, Kent and Isopi, Giovanni and Jense, Hidde T. and Keller, Ben and Kim, Joshua and Knowles, Kenda and Koopman, Brian J. and Kosowsky, Arthur and Kramer, Darby and Kusiak, Aleksandra and La Posta, Adrien and Lague, Alex and Lakey, Victoria and Lee, Eunseong and Li, Zack and Li, Yaqiong and Limon, Michele and Lokken, Martine and Louis, Thibaut and Lungu, Marius and MacCrann, Niall and MacInnis, Amanda and Maldonado, Diego and Maldonado, Felipe and Mallaby-Kay, Maya and Marques, Gabriela A. and McMahon, Jeff and Mehta, Yogesh and Menanteau, Felipe and Moodley, Kavilan and Morris, Thomas W. and Mroczkowski, Tony and Naess, Sigurd and Namikawa, Toshiya and Nati, Federico and Newburgh, Laura and Nicola, Andrina and Niemack, Michael D. and Nolta, Michael R. and Orlowski-Scherer, John and Page, Lyman A. and Pandey, Shivam and Partridge, Bruce and Prince, Heather and Puddu, Roberto and Radiconi, Federico and Robertson, Naomi and Rojas, Felipe and Sakuma, Tai and Salatino, Maria and Schaan, Emmanuel and Schmitt, Benjamin L. and Sehgal, Neelima and Shaikh, Shabbir and Sierra, Carlos and Sievers, Jon and Sifón, Cristóbal and Simon, Sara and Sonka, Rita and Spergel, David N. and Staggs, Suzanne T. and Storer, Emilie and Switzer, Eric R. and Tampier, Niklas and Thornton, Robert and Trac, Hy and Treu, Jesse and Tucker, Carole and Ullom, Joel and Vale, Leila R. and Van Engelen, Alexander and Van Lanen, Jeff and van Marrewijk, Joshiwa and Vargas, Cristian and Vavagiakis, Eve M. and Wagoner, Kasey and Wang, Yuhan and Wenzl, Lukas and Wollack, Edward J. and Xu, Zhilei and Zago, Fernando and Zheng, Kaiwen},
   year={2024},
   month=feb, pages={112} }

@article{Carron_2022_ACT+Planck,
   title={{CMB} lensing from {Planck} {PR4} maps},
   volume={2022},
   ISSN={1475-7516},
   url={http://dx.doi.org/10.1088/1475-7516/2022/09/039},
   DOI={10.1088/1475-7516/2022/09/039},
   number={09},
   journal={Journal of Cosmology and Astroparticle Physics},
   publisher={IOP Publishing},
   author={Carron, Julien and Mirmelstein, Mark and Lewis, Antony},
   year={2022},
   month=sep, pages={039} }

@ARTICLE{descollaboration2024,
       author = {{DES Collaboration} and {Abbott}, T.~M.~C. and {Acevedo}, M. and {Aguena}, M. and {Alarcon}, A. and {Allam}, S. and {Alves}, O. and {Amon}, A. and {Andrade-Oliveira}, F. and {Annis}, J. and {Armstrong}, P. and {Asorey}, J. and {Avila}, S. and {Bacon}, D. and {Bassett}, B.~A. and {Bechtol}, K. and {Bernardinelli}, P.~H. and {Bernstein}, G.~M. and {Bertin}, E. and {Blazek}, J. and {Bocquet}, S. and {Brooks}, D. and {Brout}, D. and {Buckley-Geer}, E. and {Burke}, D.~L. and {Camacho}, H. and {Camilleri}, R. and {Campos}, A. and {Carnero Rosell}, A. and {Carollo}, D. and {Carr}, A. and {Carretero}, J. and {Castander}, F.~J. and {Cawthon}, R. and {Chang}, C. and {Chen}, R. and {Choi}, A. and {Conselice}, C. and {Costanzi}, M. and {da Costa}, L.~N. and {Crocce}, M. and {Davis}, T.~M. and {DePoy}, D.~L. and {Desai}, S. and {Diehl}, H.~T. and {Dixon}, M. and {Dodelson}, S. and {Doel}, P. and {Doux}, C. and {Drlica-Wagner}, A. and {Elvin-Poole}, J. and {Everett}, S. and {Ferrero}, I. and {Fert{\'e}}, A. and {Flaugher}, B. and {Foley}, R.~J. and {Fosalba}, P. and {Friedel}, D. and {Frieman}, J. and {Frohmaier}, C. and {Galbany}, L. and {Garc{\'\i}a-Bellido}, J. and {Gatti}, M. and {Gaztanaga}, E. and {Giannini}, G. and {Glazebrook}, K. and {Graur}, O. and {Gruen}, D. and {Gruendl}, R.~A. and {Gutierrez}, G. and {Hartley}, W.~G. and {Herner}, K. and {Hinton}, S.~R. and {Hollowood}, D.~L. and {Honscheid}, K. and {Huterer}, D. and {Jain}, B. and {James}, D.~J. and {Jeffrey}, N. and {Kasai}, E. and {Kelsey}, L. and {Kent}, S. and {Kessler}, R. and {Kim}, A.~G. and {Kirshner}, R.~P. and {Kovacs}, E. and {Kuehn}, K. and {Lahav}, O. and {Lee}, J. and {Lee}, S. and {Lewis}, G.~F. and {Li}, T.~S. and {Lidman}, C. and {Lin}, H. and {Malik}, U. and {Marshall}, J.~L. and {Martini}, P. and {Mena-Fern{\'a}ndez}, J. and {Menanteau}, F. and {Miquel}, R. and {Mohr}, J.~J. and {Mould}, J. and {Muir}, J. and {M{\"o}ller}, A. and {Neilsen}, E. and {Nichol}, R.~C. and {Nugent}, P. and {Ogando}, R.~L.~C. and {Palmese}, A. and {Pan}, Y.-C. and {Paterno}, M. and {Percival}, W.~J. and {Pereira}, M.~E.~S. and {Pieres}, A. and {Malag{\'o}n}, A.~A. Plazas and {Popovic}, B. and {Porredon}, A. and {Prat}, J. and {Qu}, H. and {Raveri}, M. and {Rodr{\'\i}guez-Monroy}, M. and {Romer}, A.~K. and {Roodman}, A. and {Rose}, B. and {Sako}, M. and {Sanchez}, E. and {Sanchez Cid}, D. and {Schubnell}, M. and {Scolnic}, D. and {Sevilla-Noarbe}, I. and {Shah}, P. and {Smith}, J. Allyn. and {Smith}, M. and {Soares-Santos}, M. and {Suchyta}, E. and {Sullivan}, M. and {Suntzeff}, N. and {Swanson}, M.~E.~C. and {S{\'a}nchez}, B.~O. and {Tarle}, G. and {Taylor}, G. and {Thomas}, D. and {To}, C. and {Toy}, M. and {Troxel}, M.~A. and {Tucker}, B.~E. and {Tucker}, D.~L. and {Uddin}, S.~A. and {Vincenzi}, M. and {Walker}, A.~R. and {Weaverdyck}, N. and {Wechsler}, R.~H. and {Weller}, J. and {Wester}, W. and {Wiseman}, P. and {Yamamoto}, M. and {Yuan}, F. and {Zhang}, B. and {Zhang}, Y.},
        title = "{The Dark Energy Survey}: Cosmology Results with {\ensuremath{\sim}}1500 New High-redshift {Type Ia} {Supernovae} Using the Full 5 yr Data Set",
      journal = {\apjl},
         year = 2024,
        month = sep,
       volume = {973},
       number = {1},
          eid = {L14},
        pages = {L14},
          doi = {10.3847/2041-8213/ad6f9f},
archivePrefix = {arXiv},
       eprint = {2401.02929},
 primaryClass = {astro-ph.CO},
       adsurl = {https://ui.adsabs.harvard.edu/abs/2024ApJ...973L..14D}
}

@article{Torrado_2021_COBAYA,
   title={Cobaya: code for {Bayesian} analysis of hierarchical physical models},
   volume={2021},
   ISSN={1475-7516},
   url={http://dx.doi.org/10.1088/1475-7516/2021/05/057},
   DOI={10.1088/1475-7516/2021/05/057},
   number={05},
   journal={Journal of Cosmology and Astroparticle Physics},
   publisher={IOP Publishing},
   author={Torrado, Jesús and Lewis, Antony},
   year={2021},
   month=may, pages={057} }

@misc{2019ascl.soft10019T_COBAYA,
       author = {{Torrado}, Jes{\'u}s and {Lewis}, Antony},
        title = "{Cobaya: {Bayesian} analysis in cosmology}",
 howpublished = {Astrophysics Source Code Library, record ascl:1910.019},
         year = 2019,
        month = oct,
          eid = {ascl:1910.019},
       adsurl = {https://ui.adsabs.harvard.edu/abs/2019ascl.soft10019T}
}

@article{Lewis_2000_CAMB,
   title={Efficient Computation of {Cosmic} {Microwave} {Background} {Anisotropies} in Closed {Friedmann}‐{Robertson}‐{Walker} Models},
   volume={538},
   ISSN={1538-4357},
   url={http://dx.doi.org/10.1086/309179},
   DOI={10.1086/309179},
   number={2},
   journal={The Astrophysical Journal},
   publisher={American Astronomical Society},
   author={Lewis, Antony and Challinor, Anthony and Lasenby, Anthony},
   year={2000},
   month=aug, pages={473––476} }

@article{lewis2019_getdist,
   author = "Lewis, Antony",
   title = "{{GetDist}: a Python package for analysing {Monte Carlo} samples}",
   eprint = "1910.13970",
   archivePrefix = "arXiv",
   primaryClass = "astro-ph.IM",
   doi = "10.1088/1475-7516/2025/08/025",
   journal = "JCAP",
   volume = "08",
   pages = "025",
   year = "2025"
}

@article{Gelman_Rubin,
    author = {Andrew Gelman and Donald B. Rubin},
    title = {{Inference from Iterative Simulation Using Multiple Sequences}},
    volume = {7},
    journal = {Statistical Science},
    number = {4},
    publisher = {Institute of Mathematical Statistics},
    pages = {457 -- 472},
    year = {1992},
    doi = {10.1214/ss/1177011136},
    URL = {https://doi.org/10.1214/ss/1177011136}
}

@article{dispu_tau_able,
  title = {Addressing Tensions in $\mathrm{\ensuremath{\Lambda}}\mathrm{CDM}$ Cosmology by an Increase in the Optical Depth to Reionization},
  author = {Sailer, Noah and Farren, Gerrit S. and Ferraro, Simone and White, Martin},
  journal = {Phys. Rev. Lett.},
  volume = {136},
  issue = {8},
  pages = {081002},
  numpages = {7},
  year = {2026},
  month = {Feb},
  publisher = {American Physical Society},
  doi = {10.1103/6r54-8lv4},
  url = {https://link.aps.org/doi/10.1103/6r54-8lv4}
}

@article{DESIDR2_Neutrinos,
  title = {Constraints on neutrino physics from {DESI} {DR2} {BAO} and {DR1} full shape},
  author = {Elbers, W. and Aviles, A. and Noriega, H. E. and Chebat, D. and Menegas, A. and Frenk, C. S. and Garcia-Quintero, C. and Gonzalez, D. and Ishak, M. and Lahav, O. and Naidoo, K. and Niz, G. and Y\`eche, C. and Abdul-Karim, M. and Ahlen, S. and Alves, O. and Andrade, U. and Armengaud, E. and Behera, J. and BenZvi, S. and Bianchi, D. and Brieden, S. and Brodzeller, A. and Brooks, D. and Burtin, E. and Calderon, R. and Canning, R. and Carnero Rosell, A. and Casas, L. and Castander, F. J. and Charles, M. and Chaussidon, E. and Chaves-Montero, J. and Claybaugh, T. and Cole, S. and Cooper, A. P. and Cuceu, A. and Dawson, K. S. and de la Macorra, A. and de Mattia, A. and Deiosso, N. and Dey, A. and Dey, B. and Ding, Z. and Doel, P. and Eisenstein, D. J. and Ferraro, S. and Font-Ribera, A. and Forero-Romero, J. E. and Garrison, L. H. and Gazta\~naga, E. and Gil-Mar\'{\i}n, H. and Gontcho, S. Gontcho A. and Gonzalez-Morales, A. X. and Gutierrez, G. and He, S. and Herbold, M. and Herrera-Alcantar, H. K. and Howlett, C. and Huterer, D. and Juneau, S. and Kehoe, R. and Kirkby, D. and Kisner, T. and Kremin, A. and Lamman, C. and Landriau, M. and Le Guillou, L. and Leauthaud, A. and Levi, M. E. and Li, Q. and Lodha, K. and Magneville, C. and Manera, M. and Martini, P. and Matthewson, W. L. and Meisner, A. and Mena-Fern\'andez, J. and Miquel, R. and Moustakas, J. and Nadathur, S. and Newman, J. A. and Paillas, E. and Palanque-Delabrouille, N. and Percival, W. J. and Pieri, M. M. and Poppett, C. and Prada, F. and P\'erez-R\`afols, I. and Rabinowitz, D. and Ram\'{\i}rez-P\'erez, C. and Rashkovetskyi, M. and Ravoux, C. and Rivera-Morales, H. and Rohlf, J. and Ross, A. J. and Rossi, G. and Ruhlmann-Kleider, V. and Samushia, L. and Sanchez, E. and Schlegel, D. and Schubnell, M. and Seo, H. and Sinigaglia, F. and Sprayberry, D. and Tan, T. and Tarl\'e, G. and Taylor, P. and Turner, W. and Vargas-Maga\~na, M. and Verde, L. and Walther, M. and Weaver, B. A. and Whitford, A. and Wolfson, M. and Zarrouk, P. and Zhao, C. and Zhou, R. and Zou, H.},
  collaboration = {DESI Collaboration},
  journal = {Phys. Rev. D},
  volume = {112},
  issue = {8},
  pages = {083513},
  numpages = {33},
  year = {2025},
  month = {Oct},
  publisher = {American Physical Society},
  doi = {10.1103/w9pk-xsk7},
  url = {https://link.aps.org/doi/10.1103/w9pk-xsk7}
}

@article{itsokcurvature,
doi = {10.1088/1475-7516/2025/08/014},
url = {https://doi.org/10.1088/1475-7516/2025/08/014},
year = {2025},
month = {aug},
publisher = {IOP Publishing},
volume = {2025},
number = {08},
pages = {014},
author = {Chen, Shi-Fan and Zaldarriaga, Matias},
title = {It's all Ok: curvature in light of {BAO} from {DESI} {DR2}},
journal = {Journal of Cosmology and Astroparticle Physics}
}

@article{Neutrinos_Terrestres_1,
   title={2020 global reassessment of the neutrino oscillation picture},
   volume={2021},
   ISSN={1029-8479},
   url={http://dx.doi.org/10.1007/JHEP02(2021)071},
   DOI={10.1007/jhep02(2021)071},
   number={2},
   journal={Journal of High Energy Physics},
   publisher={Springer Science and Business Media LLC},
   author={de Salas, P. F. and Forero, D. V. and Gariazzo, S. and Martínez-Miravé, P. and Mena, O. and Ternes, C. A. and Tórtola, M. and Valle, J. W. F.},
   year={2021},
   month=feb }

@article{Neutrinos_Terrestres_2,
    author = "Capozzi, Francesco and Di Valentino, Eleonora and Lisi, Eligio and Marrone, Antonio and Melchiorri, Alessandro and Palazzo, Antonio",
    title = "{Unfinished fabric of the three neutrino paradigm}",
    eprint = "2107.00532",
    archivePrefix = "arXiv",
    primaryClass = "hep-ph",
    doi = "10.1103/PhysRevD.104.083031",
    journal = "Phys. Rev. D",
    volume = "104",
    number = "8",
    pages = "083031",
    year = "2021"
}

@article{Neutrinos_Terrestres_3,
   title={NuFit-6.0: updated global analysis of three-flavor neutrino oscillations},
   volume={2024},
   ISSN={1029-8479},
   url={http://dx.doi.org/10.1007/JHEP12(2024)216},
   DOI={10.1007/jhep12(2024)216},
   number={12},
   journal={Journal of High Energy Physics},
   publisher={Springer Science and Business Media LLC},
   author={Esteban, Ivan and Gonzalez-Garcia, M. C. and Maltoni, Michele and Martinez-Soler, Ivan and Pinheiro, João Paulo and Schwetz, Thomas},
   year={2024},
   month=dec }

@article{negativeneutrinomasses,
       author = {{Elbers}, Willem and {Frenk}, Carlos S. and {Jenkins}, Adrian and {Li}, Baojiu and {Pascoli}, Silvia},
        title = "{Negative neutrino masses as a mirage of dark energy}",
      journal = {\prd},
         year = 2025,
        month = mar,
       volume = {111},
       number = {6},
          eid = {063534},
        pages = {063534},
          doi = {10.1103/PhysRevD.111.063534},
archivePrefix = {arXiv},
       eprint = {2407.10965},
 primaryClass = {astro-ph.CO},
       adsurl = {https://ui.adsabs.harvard.edu/abs/2025PhRvD.111f3534E}
}

@article{Loverde_2024,
   title={Massive neutrinos and cosmic composition},
   volume={2024},
   ISSN={1475-7516},
   url={http://dx.doi.org/10.1088/1475-7516/2024/12/048},
   DOI={10.1088/1475-7516/2024/12/048},
   number={12},
   journal={Journal of Cosmology and Astroparticle Physics},
   publisher={IOP Publishing},
   author={Loverde, Marilena and Weiner, Zachary J.},
   year={2024},
   month=dec, pages={048} }

@article{Lesgourgues_2012,
   title={Neutrino Mass from Cosmology},
   volume={2012},
   ISSN={1687-7365},
   url={http://dx.doi.org/10.1155/2012/608515},
   DOI={10.1155/2012/608515},
   journal={Advances in High Energy Physics},
   publisher={Hindawi Limited},
   author={Lesgourgues, Julien and Pastor, Sergio},
   year={2012},
   pages={1––34} }

@article{hogg2000distancemeasurescosmology,
      title={Distance measures in cosmology}, 
      author={David W. Hogg},
      year={2000},
      eprint={astro-ph/9905116},
      archivePrefix={arXiv},
      primaryClass={astro-ph},
      url={https://arxiv.org/abs/astro-ph/9905116}, 
}

@article{CLASS_II,
   title={The Cosmic Linear Anisotropy Solving System ({CLASS}).
 {Part II}: Approximation schemes},
   volume={2011},
   ISSN={1475-7516},
   url={http://dx.doi.org/10.1088/1475-7516/2011/07/034},
   DOI={10.1088/1475-7516/2011/07/034},
   number={07},
   journal={Journal of Cosmology and Astroparticle Physics},
   publisher={IOP Publishing},
   author={Diego Blas and Julien Lesgourgues and Thomas Tram},
   year={2011},
   month=jul, pages={034––034} }

@article{CLASS_IV,
   title={The Cosmic Linear Anisotropy Solving System ({CLASS}) {IV}: efficient implementation of non-cold relics},
   volume={2011},
   ISSN={1475-7516},
   url={http://dx.doi.org/10.1088/1475-7516/2011/09/032},
   DOI={10.1088/1475-7516/2011/09/032},
   number={09},
   journal={Journal of Cosmology and Astroparticle Physics},
   publisher={IOP Publishing},
   author={Lesgourgues, Julien and Tram, Thomas},
   year={2011},
   month=sep, pages={032–032} }

@article{Gershtein1966,
    author = "Gershtein, S. S. and Zeldovich, Ya. B.",
    editor = "Srednicki, M. A.",
    title = "{Rest Mass of Muonic Neutrino and Cosmology}",
    journal = "JETP Lett.",
    volume = "4",
    pages = "120--122",
    year = "1966"
}

@article{Croft_1999,
   title={Cosmological Limits on the Neutrino Mass from the {Ly$\alpha$} Forest},
   volume={83},
   ISSN={1079-7114},
   url={http://dx.doi.org/10.1103/PhysRevLett.83.1092},
   DOI={10.1103/physrevlett.83.1092},
   number={6},
   journal={Physical Review Letters},
   publisher={American Physical Society (APS)},
   author={Croft, Rupert A. C. and Hu, Wayne and Davé, Romeel},
   year={1999},
   month=aug, pages={1092–1095} }

@ARTICLE{DESY1Anal,
       author = {{Du}, Guo-Hong and {Li}, Tian-Nuo and {Wu}, Peng-Ju and {Zhang}, Jing-Fei and {Zhang}, Xin},
        title = "{Cosmological} {Preference} for a {Positive} {Neutrino} {Mass} at 2.7$\sigma$: {A Joint Analysis of DESI DR2, DESY5, and DESY1 Data}",
      journal = {arXiv e-prints},
         year = 2025,
        month = jul,
          eid = {arXiv:2507.16589},
        pages = {arXiv:2507.16589},
          doi = {10.48550/arXiv.2507.16589},
archivePrefix = {arXiv},
       eprint = {2507.16589},
 primaryClass = {astro-ph.CO},
       adsurl = {https://ui.adsabs.harvard.edu/abs/2025arXiv250716589D}
}

@article{AA,
   title={Positive Neutrino Masses with {DESI DR2} via Matter Conversion to Dark Energy},
   volume={135},
   ISSN={1079-7114},
   DOI={10.1103/yb2k-kn7h},
   number={8},
   journal={Physical Review Letters},
   publisher={American Physical Society (APS)},
   author={Ahlen, S. P. and Aviles, A. and Cartwright, B. and Croker, K. S. and Elbers, W. and Farrah, D. and Fernandez, N. and Niz, G. and Rohlf, J. W. and Tarlé, G. and Windhorst, R. A. and Aguilar, J. and Andrade, U. and Bianchi, D. and Brooks, D. and Claybaugh, T. and de la Macorra, A. and de Mattia, A. and Dey, B. and Doel, P. and Forero-Romero, J. E. and Gaztañaga, E. and Gontcho, S. Gontcho A. and Gutierrez, G. and Huterer, D. and Ishak, M. and Kehoe, R. and Kirkby, D. and Kremin, A. and Lahav, O. and Lamman, C. and Landriau, M. and Le Guillou, L. and Levi, M. E. and Manera, M. and Miquel, R. and Moustakas, J. and Pérez-Ràfols, I. and Prada, F. and Rossi, G. and Sanchez, E. and Schubnell, M. and Seo, H. and Silber, J. and Sprayberry, D. and Walther, M. and Weaver, B. A. and Wechsler, R. H. and Zou, H.},
   year={2025},
   month=aug }

@article{axion,
  title = {Accelerated inference on accelerated cosmic expansion: New constraints on axionlike early dark energy with {DESI} {BAO} and {ACT} {DR6} {CMB} lensing},
  author = {Qu, Frank J. and Surrao, Kristen M. and Bolliet, Boris and Hill, J. Colin and Sherwin, Blake D. and Jense, Hidde T. and La Posta, Adrien},
  journal = {Phys. Rev. D},
  volume = {111},
  issue = {12},
  pages = {123507},
  numpages = {16},
  year = {2025},
  month = {Jun},
  publisher = {American Physical Society},
  doi = {10.1103/xhh6-9v62},
  url = {https://link.aps.org/doi/10.1103/xhh6-9v62}
}

@article{Helena,
doi = {10.3847/2041-8213/ae40bf},
url = {https://doi.org/10.3847/2041-8213/ae40bf},
year = {2026},
month = {feb},
publisher = {The American Astronomical Society},
volume = {998},
number = {2},
pages = {L41},
author = {García Escudero, Helena and Mirpoorian, Seyed Hamidreza and Pogosian, Levon},
title = {Sound-horizon-agnostic Inference of the {Hubble} Constant and Neutrino Masses from {Baryon Acoustic Oscillations}, {Cosmic Microwave Background} Lensing, and {Galaxy Weak} {Lensing} and {Clustering}},
journal = {The Astrophysical Journal Letters}
}

@article{Lesg,
doi = {10.1088/1475-7516/2026/02/034},
url = {https://doi.org/10.1088/1475-7516/2026/02/034},
year = {2026},
month = {feb},
publisher = {IOP Publishing},
volume = {2026},
number = {02},
pages = {034},
author = {Sharma, Ravi Kumar and Lesgourgues, Julien},
title = {Constraints on neutrino mass and dark energy agnostic to the sound horizon},
journal = {Journal of Cosmology and Astroparticle Physics}
}

@article{Liddle_2007,
   title={Information criteria for astrophysical model selection},
   volume={377},
   ISSN={1745-3925},
   url={http://dx.doi.org/10.1111/j.1745-3933.2007.00306.x},
   DOI={10.1111/j.1745-3933.2007.00306.x},
   number={1},
   journal={Monthly Notices of the Royal Astronomical Society: Letters},
   publisher={Oxford University Press (OUP)},
   author={Liddle, Andrew R.},
   year={2007},
   month=may, pages={L74–L78} }

@article{CPL2003,
   title={Exploring the Expansion History of the Universe},
   volume={90},
   ISSN={1079-7114},
   url={http://dx.doi.org/10.1103/PhysRevLett.90.091301},
   DOI={10.1103/physrevlett.90.091301},
   number={9},
   journal={Physical Review Letters},
   publisher={American Physical Society (APS)},
   author={Linder, Eric V.},
   year={2003},
   month=mar }

@article{CPL2001,
    author = "Chevallier, Michel and Polarski, David",
    title = "{Accelerating universes with scaling dark matter}",
    eprint = "gr-qc/0009008",
    archivePrefix = "arXiv",
    doi = "10.1142/S0218271801000822",
    journal = "Int. J. Mod. Phys. D",
    volume = "10",
    pages = "213--224",
    year = "2001"
}

@article{PantheonPlus,
   title={The {Pantheon+} Analysis: Cosmological Constraints},
   volume={938},
   ISSN={1538-4357},
   url={http://dx.doi.org/10.3847/1538-4357/ac8e04},
   DOI={10.3847/1538-4357/ac8e04},
   number={2},
   journal={The Astrophysical Journal},
   publisher={American Astronomical Society},
   author={Brout, Dillon and Scolnic, Dan and Popovic, Brodie and Riess, Adam G. and Carr, Anthony and Zuntz, Joe and Kessler, Rick and Davis, Tamara M. and Hinton, Samuel and Jones, David and Kenworthy, W. D’Arcy and Peterson, Erik R. and Said, Khaled and Taylor, Georgie and Ali, Noor and Armstrong, Patrick and Charvu, Pranav and Dwomoh, Arianna and Meldorf, Cole and Palmese, Antonella and Qu, Helen and Rose, Benjamin M. and Sanchez, Bruno and Stubbs, Christopher W. and Vincenzi, Maria and Wood, Charlotte M. and Brown, Peter J. and Chen, Rebecca and Chambers, Ken and Coulter, David A. and Dai, Mi and Dimitriadis, Georgios and Filippenko, Alexei V. and Foley, Ryan J. and Jha, Saurabh W. and Kelsey, Lisa and Kirshner, Robert P. and Möller, Anais and Muir, Jessie and Nadathur, Seshadri and Pan, Yen-Chen and Rest, Armin and Rojas-Bravo, Cesar and Sako, Masao and Siebert, Matthew R. and Smith, Mat and Stahl, Benjamin E. and Wiseman, Phil},
   year={2022},
   month=oct, pages={110} }

@article{Union3,
doi = {10.3847/1538-4357/adc0a5},
url = {https://doi.org/10.3847/1538-4357/adc0a5},
year = {2025},
month = {jun},
publisher = {The American Astronomical Society},
volume = {986},
number = {2},
pages = {231},
author = {Rubin, David and Aldering, Greg and Betoule, Marc and Fruchter, Andy and Huang, Xiaosheng and Kim, Alex G. and Lidman, Chris and Linder, Eric and Perlmutter, Saul and Ruiz-Lapuente, Pilar and Suzuki, Nao},
title = {Union through {UNITY}: Cosmology with 2000 {SNe} Using a Unified {Bayesian} Framework},
journal = {The Astrophysical Journal}
}

@article{DESI_DR2_II,
       author = {{Abdul Karim}, M. and {Aguilar}, J. and {Ahlen}, S. and {Alam}, S. and {Allen}, L. and {Prieto}, C. Allende and {Alves}, O. and {Anand}, A. and {Andrade}, U. and {Armengaud}, E. and {Aviles}, A. and {Bailey}, S. and {Baltay}, C. and {Bansal}, P. and {Bault}, A. and {Behera}, J. and {BenZvi}, S. and {Bianchi}, D. and {Blake}, C. and {Brieden}, S. and {Brodzeller}, A. and {Brooks}, D. and {Buckley-Geer}, E. and {Burtin}, E. and {Calderon}, R. and {Canning}, R. and {Rosell}, A. Carnero and {Carrilho}, P. and {Casas}, L. and {Castander}, F.~J. and {Charles}, M. and {Chaussidon}, E. and {Chaves-Montero}, J. and {Chebat}, D. and {Chen}, X. and {Claybaugh}, T. and {Cole}, S. and {Cooper}, A.~P. and {Cuceu}, A. and {Dawson}, K.~S. and {de la Macorra}, A. and {de Mattia}, A. and {Deiosso}, N. and {Della Costa}, J. and {Demina}, R. and {Dey}, A. and {Dey}, B. and {Ding}, Z. and {Doel}, P. and {Edelstein}, J. and {Eisenstein}, D.~J. and {Elbers}, W. and {Fagrelius}, P. and {Fanning}, K. and {Fern{\'a}ndez-Garc{\'\i}a}, E. and {Ferraro}, S. and {Font-Ribera}, A. and {Forero-Romero}, J.~E. and {Frenk}, C.~S. and {Garcia-Quintero}, C. and {Garrison}, L.~H. and {Gazta{\~n}aga}, E. and {Gil-Mar{\'\i}n}, H. and {Gontcho A Gontcho}, S. and {Gonzalez}, D. and {Gonzalez-Morales}, A.~X. and {Gordon}, C. and {Green}, D. and {Gutierrez}, G. and {Guy}, J. and {Hadzhiyska}, B. and {Hahn}, C. and {He}, S. and {Herbold}, M. and {Herrera-Alcantar}, H.~K. and {Ho}, M.-F. and {Honscheid}, K. and {Howlett}, C. and {Huterer}, D. and {Ishak}, M. and {Juneau}, S. and {Kamble}, N.~V. and {Kara{\c{c}}ayl{\i}}, N.~G. and {Kehoe}, R. and {Kent}, S. and {Kim}, A.~G. and {Kirkby}, D. and {Kisner}, T. and {Koposov}, S.~E. and {Kremin}, A. and {Krolewski}, A. and {Lahav}, O. and {Lamman}, C. and {Landriau}, M. and {Lang}, D. and {Lasker}, J. and {Le Goff}, J.~M. and {Le Guillou}, L. and {Leauthaud}, A. and {Levi}, M.~E. and {Li}, Q. and {Li}, T.~S. and {Lodha}, K. and {Lokken}, M. and {Lozano-Rodr{\'\i}guez}, F. and {Magneville}, C. and {Manera}, M. and {Martini}, P. and {Matthewson}, W.~L. and {Meisner}, A. and {Mena-Fern{\'a}ndez}, J. and {Menegas}, A. and {Mergulh{\~a}o}, T. and {Miquel}, R. and {Moustakas}, J. and {Mu{\~n}oz-Guti{\'e}rrez}, A. and {Mu{\~n}oz-Santos}, D. and {Myers}, A.~D. and {Nadathur}, S. and {Naidoo}, K. and {Napolitano}, L. and {Newman}, J.~A. and {Niz}, G. and {Noriega}, H.~E. and {Paillas}, E. and {Palanque-Delabrouille}, N. and {Pan}, J. and {Peacock}, J.~A. and {Pellejero Ibanez}, M. and {Percival}, W.~J. and {P{\'e}rez-Fern{\'a}ndez}, A. and {P{\'e}rez-R{\`a}fols}, I. and {Pieri}, M.~M. and {Poppett}, C. and {Prada}, F. and {Rabinowitz}, D. and {Raichoor}, A. and {Ram{\'\i}rez-P{\'e}rez}, C. and {Rashkovetskyi}, M. and {Ravoux}, C. and {Rich}, J. and {Rocher}, A. and {Rockosi}, C. and {Rohlf}, J. and {Rom{\'a}n-Herrera}, J.~O. and {Ross}, A.~J. and {Rossi}, G. and {Ruggeri}, R. and {Ruhlmann-Kleider}, V. and {Samushia}, L. and {Sanchez}, E. and {Sanders}, N. and {Schlegel}, D. and {Schubnell}, M. and {Seo}, H. and {Shafieloo}, A. and {Sharples}, R. and {Silber}, J. and {Sinigaglia}, F. and {Sprayberry}, D. and {Tan}, T. and {Tarl{\'e}}, G. and {Taylor}, P. and {Turner}, W. and {Ure{\~n}a-L{\'o}pez}, L.~A. and {Vaisakh}, R. and {Valdes}, F. and {Valogiannis}, G. and {Vargas-Maga{\~n}a}, M. and {Verde}, L. and {Walther}, M. and {Weaver}, B.~A. and {Weinberg}, D.~H. and {White}, M. and {Wolfson}, M. and {Y{\`e}che}, C. and {Yu}, J. and {Zaborowski}, E.~A. and {Zarrouk}, P. and {Zhai}, Z. and {Zhang}, H. and {Zhao}, C. and {Zhao}, G.~B. and {Zhou}, R. and {Zou}, H. and {DESI Collaboration}},
        title = "{{DESI} {DR2} results. {II}. Measurements of baryon acoustic oscillations and cosmological constraints}",
      journal = {\prd},
         year = 2025,
        month = oct,
       volume = {112},
       number = {8},
          eid = {083515},
        pages = {083515},
          doi = {10.1103/tr6y-kpc6},
archivePrefix = {arXiv},
       eprint = {2503.14738},
 primaryClass = {astro-ph.CO},
       adsurl = {https://ui.adsabs.harvard.edu/abs/2025PhRvD.112h3515A}
}

@article{KATRIN:2019yun,
  title = {Improved Upper Limit on the Neutrino Mass from a Direct Kinematic Method by KATRIN},
  author = {Aker, M. and Altenm\"uller, K. and Arenz, M. and Babutzka, M. and Barrett, J. and Bauer, S. and Beck, M. and Beglarian, A. and Behrens, J. and Bergmann, T. and Besserer, U. and Blaum, K. and Block, F. and Bobien, S. and Bokeloh, K. and Bonn, J. and Bornschein, B. and Bornschein, L. and Bouquet, H. and Brunst, T. and Caldwell, T. S. and La Cascio, L. and Chilingaryan, S. and Choi, W. and Corona, T. J. and Debowski, K. and Deffert, M. and Descher, M. and Doe, P. J. and Dragoun, O. and Drexlin, G. and Dunmore, J. A. and Dyba, S. and Edzards, F. and Eisenbl\"atter, L. and Eitel, K. and Ellinger, E. and Engel, R. and Enomoto, S. and Erhard, M. and Eversheim, D. and Fedkevych, M. and Felden, A. and Fischer, S. and Flatt, B. and Formaggio, J. A. and Fr\"ankle, F. M. and Franklin, G. B. and Frankrone, H. and Friedel, F. and Fuchs, D. and Fulst, A. and Furse, D. and Gauda, K. and Gemmeke, H. and Gil, W. and Gl\"uck, F. and G\"orhardt, S. and Groh, S. and Grohmann, S. and Gr\"ossle, R. and Gumbsheimer, R. and Ha Minh, M. and Hackenjos, M. and Hannen, V. and Harms, F. and Hartmann, J. and Hau\ss{}mann, N. and Heizmann, F. and Helbing, K. and Hickford, S. and Hilk, D. and Hillen, B. and Hillesheimer, D. and Hinz, D. and H\"ohn, T. and Holzapfel, B. and Holzmann, S. and Houdy, T. and Howe, M. A. and Huber, A. and James, T. M. and Jansen, A. and Kaboth, A. and Karl, C. and Kazachenko, O. and Kellerer, J. and Kernert, N. and Kippenbrock, L. and Kleesiek, M. and Klein, M. and K\"ohler, C. and K\"ollenberger, L. and Kopmann, A. and Korzeczek, M. and Kosmider, A. and Koval\'{\i}k, A. and Krasch, B. and Kraus, M. and Krause, H. and Kuckert, L. and Kuffner, B. and Kunka, N. and Lasserre, T. and Le, T. L. and Lebeda, O. and Leber, M. and Lehnert, B. and Letnev, J. and Leven, F. and Lichter, S. and Lobashev, V. M. and Lokhov, A. and Machatschek, M. and Malcherek, E. and M\"uller, K. and Mark, M. and Marsteller, A. and Martin, E. L. and Melzer, C. and Menshikov, A. and Mertens, S. and Minter, L. I. and Mirz, S. and Monreal, B. and Morales Guzm\'an, P. I. and M\"uller, K. and Naumann, U. and Ndeke, W. and Neumann, H. and Niemes, S. and Noe, M. and Oblath, N. S. and Ortjohann, H.-W. and Osipowicz, A. and Ostrick, B. and Otten, E. and Parno, D. S. and Phillips, D. G. and Plischke, P. and Pollithy, A. and Poon, A. W. P. and Pouryamout, J. and Prall, M. and Priester, F. and R\"ollig, M. and R\"ottele, C. and Ranitzsch, P. C.-O. and Rest, O. and Rinderspacher, R. and Robertson, R. G. H. and Rodenbeck, C. and Rohr, P. and Roll, Ch. and Rupp, S. and Ry\ifmmode \check{s}\else \v{s}\fi{}av\'y, M. and Sack, R. and Saenz, A. and Sch\"afer, P. and Schimpf, L. and Schl\"osser, K. and Schl\"osser, M. and Schl\"uter, L. and Sch\"on, H. and Sch\"onung, K. and Schrank, M. and Schulz, B. and Schwarz, J. and Seitz-Moskaliuk, H. and Seller, W. and Sibille, V. and Siegmann, D. and Skasyrskaya, A. and Slez\'ak, M. and \ifmmode \check{S}\else \v{S}\fi{}palek, A. and Spanier, F. and Steidl, M. and Steinbrink, N. and Sturm, M. and Suesser, M. and Sun, M. and Tcherniakhovski, D. and Telle, H. H. and Th\"ummler, T. and Thorne, L. A. and Titov, N. and Tkachev, I. and Trost, N. and Urban, K. and V\'enos, D. and Valerius, K. and VanDevender, B. A. and Vianden, R. and Vizcaya Hern\'andez, A. P. and Wall, B. L. and W\"ustling, S. and Weber, M. and Weinheimer, C. and Weiss, C. and Welte, S. and Wendel, J. and Wierman, K. J. and Wilkerson, J. F. and Wolf, J. and Xu, W. and Yen, Y.-R. and Zacher, M. and Zadorozhny, S. and Zbo\ifmmode \check{r}\else \v{r}\fi{}il, M. and Zeller, G.},
  collaboration = {KATRIN Collaboration},
  journal = {Phys. Rev. Lett.},
  volume = {123},
  issue = {22},
  pages = {221802},
  numpages = {10},
  year = {2019},
  month = {Nov},
  publisher = {American Physical Society},
  doi = {10.1103/PhysRevLett.123.221802},
  url = {https://link.aps.org/doi/10.1103/PhysRevLett.123.221802}
}

@article{KATRIN:2021uub,
    author = "Aker, M. and others",
    collaboration = "KATRIN",
    title = "{Direct neutrino-mass measurement with sub-electronvolt sensitivity}",
    eprint = "2105.08533",
    archivePrefix = "arXiv",
    primaryClass = "hep-ex",
    doi = "10.1038/s41567-021-01463-1",
    journal = "Nature Phys.",
    volume = "18",
    number = "2",
    pages = "160--166",
    year = "2022"
}

@article{KATRIN:2024cdt,
author = {KATRIN Collaboration† and Max Aker  and Dominic Batzler  and Armen Beglarian  and Jan Behrens  and Justus Beisenkötter  and Matteo Biassoni  and Benedikt Bieringer  and Yanina Biondi  and Fabian Block  and Steffen Bobien  and Matthias Böttcher  and Beate Bornschein  and Lutz Bornschein  and Tom S. Caldwell  and Marco Carminati  and Auttakit Chatrabhuti  and Suren Chilingaryan  and Byron A. Daniel  and Karol Debowski  and Martin Descher  and Deseada Díaz Barrero  and Peter J. Doe  and Otokar Dragoun  and Guido Drexlin  and Frank Edzards  and Klaus Eitel  and Enrico Ellinger  and Ralph Engel  and Sanshiro Enomoto  and Arne Felden  and Caroline Fengler  and Carlo Fiorini  and Joseph A. Formaggio  and Christian Forstner  and Florian M. Fränkle  and Kevin Gauda  and Andrew S. Gavin  and Woosik Gil  and Ferenc Glück  and Steffen Grohmann  and Robin Grössle  and Rainer Gumbsheimer  and Nathanael Gutknecht  and Volker Hannen  and Leonard Hasselmann  and Norman Haußmann  and Klaus Helbing  and Hanna Henke  and Svenja Heyns  and Stephanie Hickford  and Roman Hiller  and David Hillesheimer  and Dominic Hinz  and Thomas Höhn  and Anton Huber  and Alexander Jansen  and Christian Karl  and Jonas Kellerer  and Khanchai Khosonthongkee  and Matthias Kleifges  and Manuel Klein  and Joshua Kohpeiß  and Christoph Köhler  and Leonard Köllenberger  and Andreas Kopmann  and Neven Kovač  and Alojz Kovalík  and Holger Krause  and Luisa La Cascio  and Thierry Lasserre  and Joscha Lauer  and Thanh-Long Le  and Ondřej Lebeda  and Bjoern Lehnert  and Gen Li  and Alexey Lokhov  and Moritz Machatschek  and Martin Mark  and Alexander Marsteller  and Eric L. Martin  and Christin Melzer  and Susanne Mertens  and Shailaja Mohanty  and Jalal Mostafa  and Klaus Müller  and Andrea Nava  and Holger Neumann  and Simon Niemes  and Anthony Onillon  and Diana S. Parno  and Maura Pavan  and Udomsilp Pinsook  and Alan W. P. Poon  and Jose Manuel Lopez Poyato  and Stefano Pozzi  and Florian Priester  and Jan Ráliš  and Shivani Ramachandran  and R. G. Hamish Robertson  and Caroline Rodenbeck  and Marco Röllig  and Carsten Röttele  and Milos Ryšavý  and Rudolf Sack  and Alejandro Saenz  and Richard Salomon  and Peter Schäfer  and Magnus Schlösser  and Klaus Schlösser  and Lisa Schlüter  and Sonja Schneidewind  and Ulrich Schnurr  and Michael Schrank  and Jannis Schürmann  and Ann-Kathrin Schütz  and Alessandro Schwemmer  and Adrian Schwenck  and Michal Šefčík  and Daniel Siegmann  and Frank Simon  and Felix Spanier  and Daniela Spreng  and Warintorn Sreethawong  and Markus Steidl  and Jaroslav Štorek  and Xaver Stribl  and Michael Sturm  and Narumon Suwonjandee  and Nicholas Tan Jerome  and Helmut H. Telle  and Larisa A. Thorne  and Thomas Thümmler  and Simon Tirolf  and Nikita Titov  and Igor Tkachev  and Korbinian Urban  and Kathrin Valerius  and Drahoslav Vénos  and Christian Weinheimer  and Stefan Welte  and Jürgen Wendel  and Christoph Wiesinger  and John F. Wilkerson  and Joachim Wolf  and Sascha Wüstling  and Johanna Wydra  and Weiran Xu  and Sergey Zadorozhny  and Genrich Zeller },
title = {Direct neutrino-mass measurement based on 259 days of KATRIN data},
journal = {Science},
volume = {388},
number = {6743},
pages = {180-185},
year = {2025},
doi = {10.1126/science.adq9592},
URL = {https://www.science.org/doi/abs/10.1126/science.adq9592},
eprint = {https://www.science.org/doi/pdf/10.1126/science.adq9592}}

@article{Ma:1995ey,
    author = "Ma, Chung-Pei and Bertschinger, Edmund",
    title = "Cosmological perturbation theory in the synchronous and conformal {Newtonian} gauges",
    eprint = "astro-ph/9506072",
    archivePrefix = "arXiv",
    doi = "10.1086/176550",
    journal = "Astrophys. J.",
    volume = "455",
    pages = "7--25",
    year = "1995"
}

@article{Neutrino_FS,
  title = {Massive neutrinos in cosmology: Analytic solutions and fluid approximation},
  author = {Shoji, Masatoshi and Komatsu, Eiichiro},
  journal = {Phys. Rev. D},
  volume = {81},
  issue = {12},
  pages = {123516},
  numpages = {15},
  year = {2010},
  month = {Jun},
  publisher = {American Physical Society},
  doi = {10.1103/PhysRevD.81.123516},
  url = {https://link.aps.org/doi/10.1103/PhysRevD.81.123516}
}

@article{craig2024nusgoodnews,
    author = "Craig, Nathaniel and Green, Daniel and Meyers, Joel and Rajendran, Surjeet",
    title = "{No {\ensuremath{\nu}}s is Good News}",
    eprint = "2405.00836",
    archivePrefix = "arXiv",
    primaryClass = "astro-ph.CO",
    reportNumber = "FERMILAB-PUB-24-0492-SQMS-V",
    doi = "10.1007/JHEP09(2024)097",
    journal = "JHEP",
    volume = "09",
    pages = "097",
    year = "2024"
}

@article{naredotuero2024livingedgecriticallook,
  title = {Critical look at the cosmological neutrino mass bound},
  author = {Naredo-Tuero, Daniel and Escudero, Miguel and Enrique Fernandez-Martinez and Marcano, Xabier and Poulin, Vivian},
  journal = {Phys. Rev. D},
  volume = {110},
  issue = {12},
  pages = {123537},
  numpages = {21},
  year = {2024},
  month = {Dec},
  publisher = {American Physical Society},
  doi = {10.1103/PhysRevD.110.123537},
  url = {https://link.aps.org/doi/10.1103/PhysRevD.110.123537}
}

@article{popovic2025darkenergysurveysupernova,
    author = "Popovic, B. and others",
    collaboration = "DES",
    title = "{The Dark Energy Survey Supernova Program}: A Reanalysis Of Cosmology Results And Evidence For Evolving Dark Energy With An Updated {Type Ia Supernova Calibration}",
    eprint = "2511.07517",
    archivePrefix = "arXiv",
    primaryClass = "astro-ph.CO",
    reportNumber = "FERMILAB-PUB-25-0842-CSAID-PPD",
    month = "11",
    year = "2025"
}

@article{Planck:2013nga,
    author = "Ade, P. A. R. and others",
    collaboration = "Planck",
    title = "{Planck intermediate results. XVI. Profile likelihoods for cosmological parameters}",
    eprint = "1311.1657",
    archivePrefix = "arXiv",
    primaryClass = "astro-ph.CO",
    doi = "10.1051/0004-6361/201323003",
    journal = "Astron. Astrophys.",
    volume = "566",
    pages = "A54",
    year = "2014"
}

@article{Palanque_Delabrouille_2015,
   title={Neutrino masses and cosmology with {Lyman}-alpha forest power spectrum},
   volume={2015},
   ISSN={1475-7516},
   url={http://dx.doi.org/10.1088/1475-7516/2015/11/011},
   DOI={10.1088/1475-7516/2015/11/011},
   number={11},
   journal={Journal of Cosmology and Astroparticle Physics},
   publisher={IOP Publishing},
   author={Palanque-Delabrouille, Nathalie and Yèche, Christophe and Baur, Julien and Magneville, Christophe and Rossi, Graziano and Lesgourgues, Julien and Borde, Arnaud and Burtin, Etienne and LeGoff, Jean-Marc and Rich, James and Viel, Matteo and Weinberg, David},
   year={2015},
   month=nov, pages={011–011} }

@article{Alam_2021,
   title={Completed {SDSS-IV} extended {Baryon} {Oscillation} {Spectroscopic} {Survey}: Cosmological implications from two decades of spectroscopic surveys at the {Apache} {Point} {Observatory}},
   volume={103},
   ISSN={2470-0029},
   url={http://dx.doi.org/10.1103/PhysRevD.103.083533},
   DOI={10.1103/physrevd.103.083533},
   number={8},
   journal={Physical Review D},
   publisher={American Physical Society (APS)},
   author={Alam, Shadab and Aubert, Marie and Avila, Santiago and Balland, Christophe and Bautista, Julian E. and Bershady, Matthew A. and Bizyaev, Dmitry and Blanton, Michael R. and Bolton, Adam S. and Bovy, Jo and Brinkmann, Jonathan and Brownstein, Joel R. and Burtin, Etienne and Chabanier, Solène and Chapman, Michael J. and Choi, Peter Doohyun and Chuang, Chia-Hsun and Comparat, Johan and Cousinou, Marie-Claude and Cuceu, Andrei and Dawson, Kyle S. and de la Torre, Sylvain and de Mattia, Arnaud and Agathe, Victoria de Sainte and des Bourboux, Hélion du Mas and Escoffier, Stephanie and Etourneau, Thomas and Farr, James and Font-Ribera, Andreu and Frinchaboy, Peter M. and Fromenteau, Sebastien and Gil-Marín, Héctor and Le Goff, Jean-Marc and Gonzalez-Morales, Alma X. and Gonzalez-Perez, Violeta and Grabowski, Kathleen and Guy, Julien and Hawken, Adam J. and Hou, Jiamin and Kong, Hui and Parker, James and Klaene, Mark and Kneib, Jean-Paul and Lin, Sicheng and Long, Daniel and Lyke, Brad W. and de la Macorra, Axel and Martini, Paul and Masters, Karen and Mohammad, Faizan G. and Moon, Jeongin and Mueller, Eva-Maria and Muñoz-Gutiérrez, Andrea and Myers, Adam D. and Nadathur, Seshadri and Neveux, Richard and Newman, Jeffrey A. and Noterdaeme, Pasquier and Oravetz, Audrey and Oravetz, Daniel and Palanque-Delabrouille, Nathalie and Pan, Kaike and Paviot, Romain and Percival, Will J. and Pérez-Ràfols, Ignasi and Petitjean, Patrick and Pieri, Matthew M. and Prakash, Abhishek and Raichoor, Anand and Ravoux, Corentin and Rezaie, Mehdi and Rich, James and Ross, Ashley J. and Rossi, Graziano and Ruggeri, Rossana and Ruhlmann-Kleider, Vanina and Sánchez, Ariel G. and Sánchez, F. Javier and Sánchez-Gallego, José R. and Sayres, Conor and Schneider, Donald P. and Seo, Hee-Jong and Shafieloo, Arman and Slosar, Anže and Smith, Alex and Stermer, Julianna and Tamone, Amelie and Tinker, Jeremy L. and Tojeiro, Rita and Vargas-Magaña, Mariana and Variu, Andrei and Wang, Yuting and Weaver, Benjamin A. and Weijmans, Anne-Marie and Yèche, Christophe and Zarrouk, Pauline and Zhao, Cheng and Zhao, Gong-Bo and Zheng, Zheng},
   year={2021},
   month=apr }

@article{Aiola_2020,
   title={The {Atacama} {Cosmology} {Telescope}: {DR4} maps and cosmological parameters},
   volume={2020},
   ISSN={1475-7516},
   url={http://dx.doi.org/10.1088/1475-7516/2020/12/047},
   DOI={10.1088/1475-7516/2020/12/047},
   number={12},
   journal={Journal of Cosmology and Astroparticle Physics},
   publisher={IOP Publishing},
   author={Aiola, Simone and Calabrese, Erminia and Maurin, Loïc and Naess, Sigurd and Schmitt, Benjamin L. and Abitbol, Maximilian H. and Addison, Graeme E. and Ade, Peter A. R. and Alonso, David and Amiri, Mandana and Amodeo, Stefania and Angile, Elio and Austermann, Jason E. and Baildon, Taylor and Battaglia, Nick and Beall, James A. and Bean, Rachel and Becker, Daniel T. and Bond, J Richard and Bruno, Sarah Marie and Calafut, Victoria and Campusano, Luis E. and Carrero, Felipe and Chesmore, Grace E. and Cho, Hsiao-mei and Choi, Steve K. and Clark, Susan E. and Cothard, Nicholas F. and Crichton, Devin and Crowley, Kevin T. and Darwish, Omar and Datta, Rahul and Denison, Edward V. and Devlin, Mark J. and Duell, Cody J. and Duff, Shannon M. and Duivenvoorden, Adriaan J. and Dunkley, Jo and Dünner, Rolando and Essinger-Hileman, Thomas and Fankhanel, Max and Ferraro, Simone and Fox, Anna E. and Fuzia, Brittany and Gallardo, Patricio A. and Gluscevic, Vera and Golec, Joseph E. and Grace, Emily and Gralla, Megan and Guan, Yilun and Hall, Kirsten and Halpern, Mark and Han, Dongwon and Hargrave, Peter and Hasselfield, Matthew and Helton, Jakob M. and Henderson, Shawn and Hensley, Brandon and Hill, J. Colin and Hilton, Gene C. and Hilton, Matt and Hincks, Adam D. and Hložek, Renée and Ho, Shuay-Pwu Patty and Hubmayr, Johannes and Huffenberger, Kevin M. and Hughes, John P. and Infante, Leopoldo and Irwin, Kent and Jackson, Rebecca and Klein, Jeff and Knowles, Kenda and Koopman, Brian and Kosowsky, Arthur and Lakey, Vincent and Li, Dale and Li, Yaqiong and Li, Zack and Lokken, Martine and Louis, Thibaut and Lungu, Marius and MacInnis, Amanda and Madhavacheril, Mathew and Maldonado, Felipe and Mallaby-Kay, Maya and Marsden, Danica and McMahon, Jeff and Menanteau, Felipe and Moodley, Kavilan and Morton, Tim and Namikawa, Toshiya and Nati, Federico and Newburgh, Laura and Nibarger, John P. and Nicola, Andrina and Niemack, Michael D. and Nolta, Michael R. and Orlowski-Sherer, John and Page, Lyman A. and Pappas, Christine G. and Partridge, Bruce and Phakathi, Phumlani and Pisano, Giampaolo and Prince, Heather and Puddu, Roberto and Qu, Frank J. and Rivera, Jesus and Robertson, Naomi and Rojas, Felipe and Salatino, Maria and Schaan, Emmanuel and Schillaci, Alessandro and Sehgal, Neelima and Sherwin, Blake D. and Sierra, Carlos and Sievers, Jon and Sifon, Cristobal and Sikhosana, Precious and Simon, Sara and Spergel, David N. and Staggs, Suzanne T. and Stevens, Jason and Storer, Emilie and Sunder, Dhaneshwar D. and Switzer, Eric R. and Thorne, Ben and Thornton, Robert and Trac, Hy and Treu, Jesse and Tucker, Carole and Vale, Leila R. and Engelen, Alexander Van and Lanen, Jeff Van and Vavagiakis, Eve M. and Wagoner, Kasey and Wang, Yuhan and Ward, Jonathan T. and Wollack, Edward J. and Xu, Zhilei and Zago, Fernando and Zhu, Ningfeng},
   year={2020},
   month=dec, pages={047–047} }

@article{Di_Valentino_2022,
   title={Health checkup test of the standard cosmological model in view of recent cosmic microwave background anisotropies experiments},
   volume={106},
   ISSN={2470-0029},
   url={http://dx.doi.org/10.1103/PhysRevD.106.103506},
   DOI={10.1103/physrevd.106.103506},
   number={10},
   journal={Physical Review D},
   publisher={American Physical Society (APS)},
   author={Di Valentino, Eleonora and Giarè, William and Melchiorri, Alessandro and Silk, Joseph},
   year={2022},
   month=nov }

@article{Chudaykin:2026nls,
    author = "Chudaykin, Anton and Ivanov, Mikhail M. and Philcox, Oliver H. E.",
    title = "Reanalyzing {DESI} {DR1}: 5. {Cosmological Constraints with Simulation-Based Priors}",
    eprint = "2602.18554",
    archivePrefix = "arXiv",
    primaryClass = "astro-ph.CO",
    reportNumber = "MIT-CTP/6007",
    month = "2",
    year = "2026"
}

@article{DESI:2025fii,
    author = "Lodha, K. and others",
    collaboration = "DESI",
    title = "Extended dark energy analysis using {DESI} {DR2} {BAO} measurements",
    eprint = "2503.14743",
    archivePrefix = "arXiv",
    primaryClass = "astro-ph.CO",
    reportNumber = "FERMILAB-PUB-25-0164-PPD",
    doi = "10.1103/w4c6-1r5j",
    journal = "Phys. Rev. D",
    volume = "112",
    number = "8",
    pages = "083511",
    year = "2025"
}

@article{DESI:2025wyn,
    author = "Gu, Gan and others",
    collaboration = "DESI",
    title = "Dynamical dark energy in light of the {DESI} {DR2} baryonic acoustic oscillations measurements",
    eprint = "2504.06118",
    archivePrefix = "arXiv",
    primaryClass = "astro-ph.CO",
    reportNumber = "FERMILAB-PUB-25-0235-PPD",
    doi = "10.1038/s41550-025-02669-6",
    journal = "Nature Astron.",
    volume = "9",
    number = "12",
    pages = "1879--1889",
    year = "2025",
    note = "[Erratum: Nature Astron. 9, 1898 (2025)]"
}

@article{Calabrese_2025,
doi = {10.1088/1475-7516/2025/11/063},
url = {https://doi.org/10.1088/1475-7516/2025/11/063},
year = {2025},
month = {nov},
publisher = {IOP Publishing},
volume = {2025},
number = {11},
pages = {063},
author = {Calabrese, Erminia and Hill, J. Colin and Jense, Hidde T. and La Posta, Adrien and Abril-Cabezas, Irene and Addison, Graeme E. and Ade, Peter A.R. and Aiola, Simone and Alford, Tommy and Alonso, David and Amiri, Mandana and An, Rui and Atkins, Zachary and Austermann, Jason E. and Barbavara, Eleonora and Barbieri, Nicola and Battaglia, Nicholas and Battistelli, Elia Stefano and Beall, James A. and Bean, Rachel and Beheshti, Ali and Beringue, Benjamin and Bhandarkar, Tanay and Biermann, Emily and Bolliet, Boris and Bond, J Richard and Capalbo, Valentina and Carrero, Felipe and Chen, Shi-Fan and Chesmore, Grace and Cho, Hsiao-mei and Choi, Steve K. and Clark, Susan E. and Cothard, Nicholas F. and Coughlin, Kevin and Coulton, William and Crichton, Devin and Crowley, Kevin T. and Darwish, Omar and Devlin, Mark J. and Dicker, Simon and Duell, Cody J. and Duff, Shannon M. and Duivenvoorden, Adriaan J. and Dunkley, Jo and Dunner, Rolando and Embil Villagra, Carmen and Fankhanel, Max and Farren, Gerrit S. and Ferraro, Simone and Foster, Allen and Freundt, Rodrigo and Fuzia, Brittany and Gallardo, Patricio A. and Garrido, Xavier and Gerbino, Martina and Giardiello, Serena and Gill, Ajay and Givans, Jahmour and Gluscevic, Vera and Goldstein, Samuel and Golec, Joseph E. and Gong, Yulin and Guan, Yilun and Halpern, Mark and Harrison, Ian and Hasselfield, Matthew and He, Adam and Healy, Erin and Henderson, Shawn and Hensley, Brandon and Hervías-Caimapo, Carlos and Hilton, Gene C. and Hilton, Matt and Hincks, Adam D. and Hložek, Renée and Ho, Shuay-Pwu Patty and Hood, John and Hornecker, Erika and Huber, Zachary B. and Hubmayr, Johannes and Huffenberger, Kevin M. and Hughes, John P. and Ikape, Margaret and Irwin, Kent and Isopi, Giovanni and Joshi, Neha and Keller, Ben and Kim, Joshua and Knowles, Kenda and Koopman, Brian J. and Kosowsky, Arthur and Kramer, Darby and Kusiak, Aleksandra and Laguë, Alex and Lakey, Victoria and Lattanzi, Massimiliano and Lee, Eunseong and Li, Yaqiong and Li, Zack and Limon, Michele and Lokken, Martine and Louis, Thibaut and Lungu, Marius and MacCrann, Niall and MacInnis, Amanda and Madhavacheril, Mathew S. and Maldonado, Diego and Maldonado, Felipe and Mallaby-Kay, Maya and Marques, Gabriela A. and van Marrewijk, Joshiwa and McCarthy, Fiona and McMahon, Jeff and Mehta, Yogesh and Menanteau, Felipe and Moodley, Kavilan and Morris, Thomas W. and Mroczkowski, Tony and Naess, Sigurd and Namikawa, Toshiya and Nati, Federico and Nerval, Simran K. and Newburgh, Laura and Nicola, Andrina and Niemack, Michael D. and Nolta, Michael R. and Orlowski-Scherer, John and Pagano, Luca and Page, Lyman A. and Pandey, Shivam and Partridge, Bruce and Perez Sarmiento, Karen and Prince, Heather and Puddu, Roberto and Qu, Frank J. and Ragavan, Damien C. and Ried Guachalla, Bernardita and Rogers, Keir K. and Rojas, Felipe and Sakuma, Tai and Schaan, Emmanuel and Schmitt, Benjamin L. and Sehgal, Neelima and Shaikh, Shabbir and Sherwin, Blake D. and Sierra, Carlos and Sievers, Jon and Sifón, Cristóbal and Simon, Sara and Sonka, Rita and Spergel, David N. and Staggs, Suzanne T. and Storer, Emilie and Surrao, Kristen and Switzer, Eric R. and Tampier, Niklas and Thiele, Leander and Thornton, Robert and Trac, Hy and Tucker, Carole and Ullom, Joel and Vale, Leila R. and Van Engelen, Alexander and Van Lanen, Jeff and Vargas, Cristian and Vavagiakis, Eve M. and Wagoner, Kasey and Wang, Yuhan and Wenzl, Lukas and Wollack, Edward J. and Zheng, Kaiwen and The Atacama Cosmology Telescope collaboration},
title = {The {Atacama} {Cosmology} {Telescope}: {DR6} constraints on extended cosmological models},
journal = {Journal of Cosmology and Astroparticle Physics}
}

@article{Jense_2026,
   title={Choice of Planck CMB likelihood in cosmological analyses},
   volume={113},
   ISSN={2470-0029},
   url={http://dx.doi.org/10.1103/ff9j-skyj},
   DOI={10.1103/ff9j-skyj},
   number={4},
   journal={Physical Review D},
   publisher={American Physical Society (APS)},
   author={Jense, Hidde T. and Viña, Marc and Calabrese, Erminia and Hill, J. Colin},
   year={2026},
   month=feb }

@article{Challenging-LCDM,
  title = {Challenging the $\mathrm{\ensuremath{\Lambda}}\mathrm{CDM}$ model: $5\ensuremath{\sigma}$ evidence for a dynamical dark energy late-time transition},
  author = {Scherer, Mateus and Sabogal, Miguel A. and Nunes, Rafael C. and De Felice, Antonio},
  journal = {Phys. Rev. D},
  volume = {112},
  issue = {4},
  pages = {043513},
  numpages = {15},
  year = {2025},
  month = {Aug},
  publisher = {American Physical Society},
  doi = {10.1103/n86r-sjgm},
  url = {https://link.aps.org/doi/10.1103/n86r-sjgm}
}

@article{101093/mnrasl/slaf063,
    author = {Dinda, Bikash R and Maartens, Roy},
    title = {Physical versus phantom dark energy after DESI: thawing quintessence in a curved background},
    journal = {Monthly Notices of the Royal Astronomical Society: Letters},
    volume = {542},
    number = {1},
    pages = {L31-L35},
    year = {2025},
    month = {09},
    issn = {1745-3925},
    doi = {10.1093/mnrasl/slaf063},
    url = {https://doi.org/10.1093/mnrasl/slaf063},
    eprint = {https://academic.oup.com/mnrasl/article-pdf/542/1/L31/63485376/slaf063.pdf},
}

@article{q1ng-2xvp,
  title = {Joint constraints on neutrinos and dynamical dark energy in minimally modified gravity},
  author = {Ladeira, Artur and Nunes, Rafael C. and Pan, Supriya and Yang, Weiqiang},
  journal = {Phys. Rev. D},
  volume = {113},
  issue = {8},
  pages = {083503},
  numpages = {19},
  year = {2026},
  month = {Apr},
  publisher = {American Physical Society},
  doi = {10.1103/q1ng-2xvp},
  url = {https://link.aps.org/doi/10.1103/q1ng-2xvp}
}

@article{lwdv-qrcc,
  title = {Neutrino decays as a natural explanation of the neutrino mass tension},
  author = {Franco Abell\'an, Guillermo},
  journal = {Phys. Rev. D},
  volume = {113},
  issue = {12},
  pages = {123527},
  numpages = {12},
  year = {2026},
  month = {Jun},
  publisher = {American Physical Society},
  doi = {10.1103/lwdv-qrcc},
  url = {https://link.aps.org/doi/10.1103/lwdv-qrcc}
}

@article{Roy_Choudhury_2025,
   title={Cosmology in Extended Parameter Space with DESI Data Release 2 Baryon Acoustic Oscillations: A 2σ+ Detection of Nonzero Neutrino Masses with an Update on Dynamical Dark Energy and Lensing Anomaly},
   volume={986},
   ISSN={2041-8213},
   url={http://dx.doi.org/10.3847/2041-8213/ade1cc},
   DOI={10.3847/2041-8213/ade1cc},
   number={2},
   journal={The Astrophysical Journal Letters},
   publisher={American Astronomical Society},
   author={Roy Choudhury, Shouvik},
   year={2025},
   month=jun, pages={L31} }

@article{Trotta_2008,
   title={Bayes in the sky: Bayesian inference and model selection in cosmology},
   volume={49},
   ISSN={1366-5812},
   url={http://dx.doi.org/10.1080/00107510802066753},
   DOI={10.1080/00107510802066753},
   number={2},
   journal={Contemporary Physics},
   publisher={Informa UK Limited},
   author={Trotta, Roberto},
   year={2008},
   month=Mar, pages={71–104} }

@misc{mcewen2023machinelearningassistedbayesian,
      title={Machine learning assisted Bayesian model comparison: \\ learnt harmonic mean estimator}, 
      author={Jason D. McEwen and Christopher G. R. Wallis and Matthew A. Price and Alessio Spurio Mancini},
      year={2023},
      eprint={2111.12720},
      archivePrefix={arXiv},
      primaryClass={stat.ME},
      url={https://arxiv.org/abs/2111.12720}, 
}

\newpage
\appendix

\section{Dovekie results}\label{Apendix: Dovekie_Results}

\begin{figure}[!h]
    \centering
    \includegraphics[width=1 \linewidth]{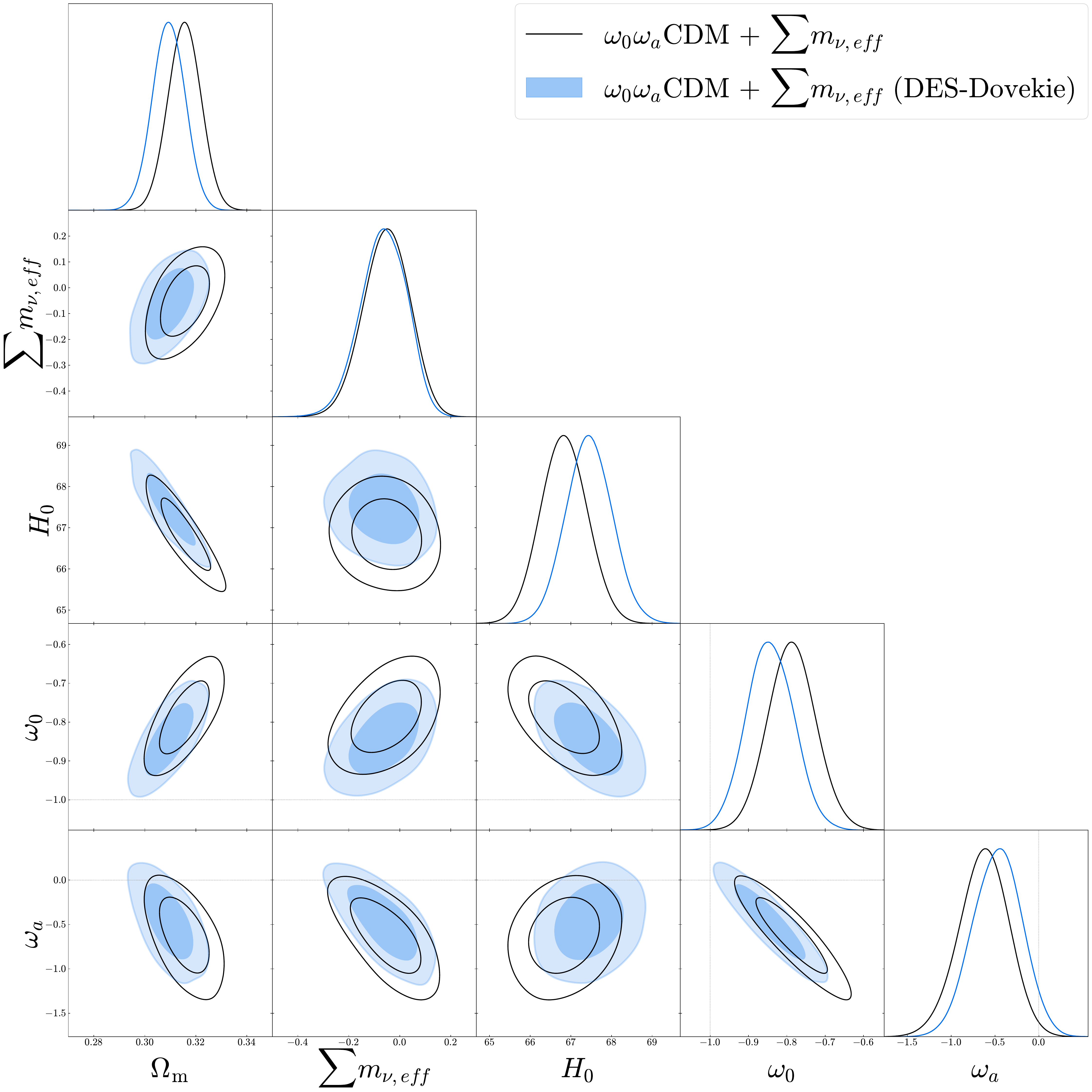}
    \caption{Comparison of the $\omega_0\omega_a$CDM + $\sum m_{\nu,eff}$ model between DESY5 and DESY5-Dovekie datasets. The reanalysis impacts several parameters, yet the neutrino mass posterior is invariant.}
    \label{fig:Flat_Dove}
\end{figure}

\begin{figure}[p]
    \centering
    \includegraphics[width=1 \linewidth]{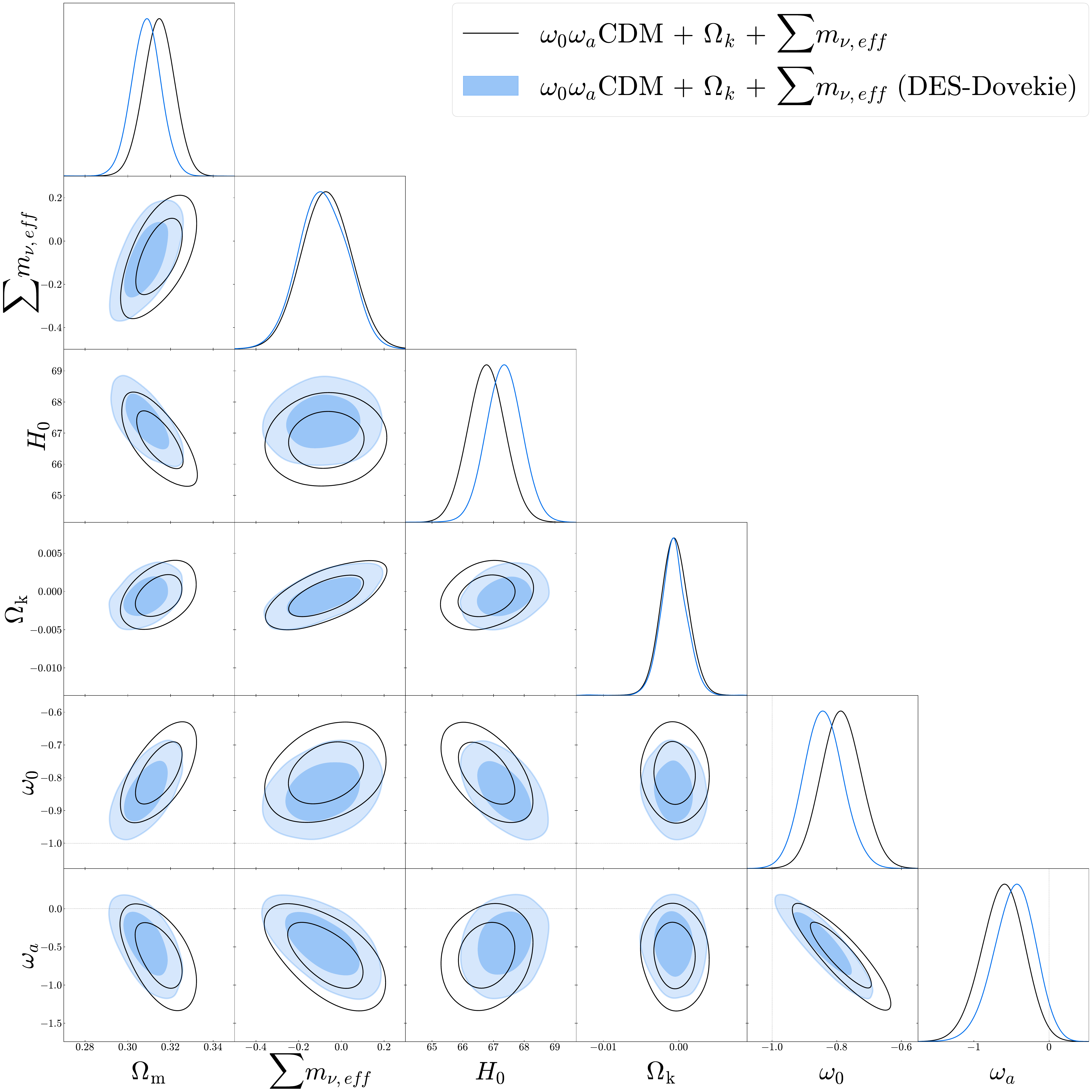}
    \caption{Comparison of the $\omega_0\omega_a$CDM + $\Omega_k$ + $\sum m_{\nu,eff}$ model between DESY5 and DESY5-Dovekie datasets. As in Figure \ref{fig:Flat_Dove}, the $\sum m_{\nu,eff}$ distribution is unaffected by the reanalysis, and $\Omega_k$ exhibits a similar behavior.}
    \label{fig:Curved_Dove}
\end{figure}

A recent reanalysis of the DES Supernova dataset, designated as DESY5-Dovekie \cite{popovic2025darkenergysurveysupernova}, has been released. These updated data reduce the tension between $\Lambda$CDM and $\omega_0\omega_a$CDM from $4.2\sigma$ to $3.2\sigma$, thereby mitigating the shift towards an evolving dark energy scenario. Given these findings, and considering the dataset tensions discussed in Section \ref{Sec:Datasets}, it was necessary to investigate the $\omega_0\omega_a$CDM scenarios with negative neutrino masses using this new dataset to assess any emerging shifts. Figures \ref{fig:Flat_Dove} and \ref{fig:Curved_Dove} show the resulting posteriors, comparing the original DESY5 with DESY5-Dovekie. 

The implementation of the DES-Dovekie reanalysis demonstrates that the posterior distribution 
for $\sum m_{\nu,eff}$ remains essentially invariant, even when other parameters, 
like $\Omega_m$, exhibit noticeable shifts. The fact that the neutrino mass sum is 
preserved under this updated calibration reinforces the robustness of our main findings, 
confirming that our primary results are insensitive to this specific treatment.

\clearpage
\onecolumngrid
\section{Best-fit results for all cosmological parameters}\label{Apendix: All_Results}

To maintain clarity in the main discussion, the detailed results for the extended parameter set have been moved to this appendix.

\begin{table}[!h]
\centering
\makebox[\textwidth][c]{%
    \resizebox{1\textwidth}{!}{%
\begin{tabular}{l c c c c c c c c c c c c}
\toprule
\textbf{Model} & $\mathbf{\Omega_m}$ & $\mathbf{\Omega_{cdm}h^2}$ & $\mathbf{\Omega_{b}h^2}$ & $\mathbf{H_0}$ [km s$\mathbf{^{-1}}$ Mpc$\mathbf{^{-1}}$] & $100\mathbf{\theta_{*}}$ & $\ln{\left(10^{-9}\mathbf{A_s}\right)}$ & $\mathbf{n_s}$ & $\mathbf{\tau}$ & $\mathbf{10^3 \Omega_k}$ & $\mathbf{\sum m_\nu}$ [eV] & $\mathbf{\omega_0}$ & $\mathbf{\omega_a}$ \\
\midrule
$\mathbf{\Lambda}$CDM  & $0.3049\pm 0.0036$ & $0.11814\pm 0.00061$ & $0.02230\pm 0.00012$ & $68.03\pm 0.27$ & $1.04096\pm 0.00023$ & $3.051\pm 0.013$ & $0.9681\pm 0.0033$ & $0.0591^{+0.0066}_{-0.0073}$ & - & - & - & -  \\
$\mathbf{\Lambda}$CDM + $\mathbf{\sum m_\nu}$ & $0.3035\pm 0.0036$ & $0.11841\pm 0.00064$ & $0.02229\pm 0.00012$ & $68.17\pm 0.29$ & $1.04094\pm 0.00023$ & $3.047\pm 0.013$ & $0.9673\pm 0.0034$ & $0.0572\pm 0.0070$ & - & $< 0.0731$ & - & -  \\
$\mathbf{\Lambda}$CDM + $\mathbf{\Omega_k}$ & $0.3058\pm 0.0036$ & $0.1200\pm 0.0011$ & $0.02218\pm 0.00014$ & $68.35\pm 0.32$ & $1.04075\pm 0.00026$ & $3.048^{+0.011}_{-0.013}$ & $0.9637\pm 0.0039$ & $0.0558^{+0.0064}_{-0.0073}$ & $2.5\pm 1.2$ & - & - & -  \\
$\mathbf{\Lambda}$CDM + $\mathbf{\Omega_k}$ + $\mathbf{\sum m_\nu}$ & $0.3055^{+0.0039}_{-0.0044}$ & $0.1200\pm 0.0011$ & $0.02218\pm 0.00014$ & $68.36\pm 0.34$ & $1.04074\pm 0.00026$ & $3.047\pm 0.013$ & $0.9636\pm 0.0041$ & $0.0556\pm 0.0072$ & $2.3^{+1.2}_{-1.4}$ & $< 0.135$ & - & -  \\
$\mathbf{\omega_0\omega_a}$CDM + $\mathbf{\sum m_{\nu}}$ & $0.3187\pm 0.0057$ & $0.11919\pm 0.00083$ & $0.02224\pm 0.00013$ & $66.75\pm 0.56$ & $1.04082\pm 0.00024$ & $3.041\pm 0.013$ & $0.9654\pm 0.0036$ & $0.0539\pm 0.0072$ & - & $< 0.130$ & -$0.758\pm 0.059$ & -$0.82^{+0.26}_{-0.21}$  \\
$\mathbf{\omega_0\omega_a}$CDM + $\mathbf{\Omega_k}$ & $0.3187\pm 0.0059$ & $0.1196\pm 0.0011$ & $0.02222\pm 0.00014$ & $66.87\pm 0.62$ & $1.04079\pm 0.00026$ & $3.042\pm 0.013$ & $0.9645\pm 0.0040$ & $0.0537\pm 0.0071$ & $0.9\pm 1.4$ & - & -$0.774^{+0.073}_{-0.057}$ & -$0.76^{+0.23}_{-0.29}$  \\
$\mathbf{\omega_0\omega_a}$CDM + $\mathbf{\Omega_k}$ + $\mathbf{\sum m_\nu}$ & $0.3190\pm 0.0061$ & $0.1196\pm 0.0012$ & $0.02222\pm 0.00014$ & $66.85\pm 0.61$ & $1.04078\pm 0.00027$ & $3.042\pm 0.014$ & $0.9647\pm 0.0042$ & $0.0538\pm 0.0074$ & $0.9^{+1.4}_{-1.6}$ & $< 0.171$ & -$0.773^{+0.074}_{-0.060}$ & -$0.76^{+0.26}_{-0.30}$  \\
\midrule
$\mathbf{\Lambda}$CDM + $\mathbf{\sum m_{\nu,eff}}$ & $0.2992\pm 0.0045$ & $0.11928\pm 0.00083$ & $0.02221\pm 0.00013$ & $68.54\pm 0.37$ & $1.04177\pm 0.00023$ & $3.036\pm 0.014$ & $0.9645\pm 0.0037$ & $0.0521\pm 0.0076$ & - & -$0.073^{+0.051}_{-0.064}$ & - & -  \\
$\mathbf{\Lambda}$CDM + $\mathbf{\Omega_k}$ + $\mathbf{\sum m_{\nu,eff}}$ & $0.3013\pm 0.0054$ & $0.1198\pm 0.0011$ & $0.02218\pm 0.00014$ & $68.55\pm 0.39$ & $1.04173\pm 0.00024$ & $3.038\pm 0.015$ & $0.9633\pm 0.0040$ & $0.0521\pm 0.0077$ & $1.1^{+1.5}_{-1.7}$ & -$0.011^{+0.052}_{-0.050}$ & - & -  \\
$\mathbf{\omega_0\omega_a}$CDM + $\mathbf{\sum m_{\nu,eff}}$ & $0.3156\pm 0.0063$ & $0.11944\pm 0.00087$ & $0.02221^{+0.00013}_{-0.00014}$ & $66.83\pm 0.56$ & $1.04175\pm 0.00024$ & $3.034\pm 0.015$ & $0.9640\pm 0.0038$ & $0.0509\pm 0.0076$ & - & -$0.052^{+0.095}_{-0.084}$ & -$0.787\pm 0.061$ & -$0.63^{+0.29}_{-0.26}$  \\
$\mathbf{\omega_0\omega_a}$CDM + $\mathbf{\Omega_k}$ + $\mathbf{\sum m_{\nu,eff}}$ & $0.3146\pm 0.0071$ & $0.1191\pm 0.0012$ & $0.02225\pm 0.00014$ & $66.78\pm 0.60$ & $1.04177\pm 0.00025$ & $3.031\pm 0.016$ & $0.9652\pm 0.0043$ & $0.0503\pm 0.0079$ & -$0.6^{+1.6}_{-1.9}$ & -$0.07\pm 0.11$ & -$0.786\pm 0.061$ & -$0.61^{+0.29}_{-0.26}$  \\
\bottomrule
\end{tabular}%
}
}
\caption{Cosmological parameter constraints for all analyzed models. All values are reported at 68\% confidence level, except for $\sum m_\nu$, which corresponds to the 95\% upper limit. }
\label{tab:Results_all}
\end{table}

\end{document}